\documentclass[aps,floatfix,eqsecnum,showpacs,preprint,tightenlines]{revtex4}
\usepackage{epsfig}
\usepackage{amsmath}

\def\bsigma{\mbox{\boldmath$\sigma$}}
\def\btau{\mbox{\boldmath$\tau$}}
\def\e{\kern+.6ex\lower.42ex\hbox{$\scriptstyle \iota$}\kern-1.20ex e}
\begin{document}

\title{
The Two-Nucleon System in Three Dimensions
}
 
\author{J.~Golak$^1$}

\author{W.~Gl\"ockle$^2$}

\author{R.~Skibi\'nski$^1$}

\author{H.~Wita{\l}a$^1$}

\author{D.~Rozp{\e}dzik$^1$}

\author{K.~Topolnicki$^1$}

\author{I.~Fachruddin$^3$}

\author{Ch.~Elster$^4$}

\author{A.~Nogga$^{5}$}

\affiliation{$^1$M. Smoluchowski Institute of Physics, Jagiellonian
University, PL-30059 Krak\'ow, Poland}

\affiliation{$^2$Institut f\"ur Theoretische Physik II,
Ruhr-Universit\"at Bochum, D-44780 Bochum, Germany}

\affiliation{$^3$Departemen Fisika, Universitas Indonesia, 
Depok 16424, Indonesia}

\affiliation{$^4$Institute of Nuclear and Particle Physics, 
Department of Physics and Astronomy, Ohio University, Athens, OH 45701, USA}

\affiliation{$^5$Forschungszentrum J\"ulich,
          Institut f\"ur Kernphysik (Theorie), 
          Institute for Advanced Simulation
          and  J\"ulich Center for Hadron Physics, D-52425 J\"ulich, Germany}

\date{\today}

\begin{abstract}
A recently developed formulation for treating two- and three-nucleon
bound states in a three-dimensional formulation based on spin-momentum operators is
extended to nucleon-nucleon scattering. 
Here the nucleon-nucleon t-matrix is represented by six spin-momentum
operators
accompanied by six scalar functions of momentum vectors. We present
the formulation and  provide numerical
examples for the deuteron and nucleon-nucleon
scattering observables. A comparison to results from a standard
partial wave decomposition establishes the reliability of this new
formulation.
\end{abstract}
\maketitle \setcounter{page}{1}

\section{Introduction}

A standard way to obtain scattering observables for nucleon-nucleon (NN) scattering
is to solve the Schr\"odinger equation either in momentum or coordinate space 
by taking advantage of rotational invariance and introduce a partial wave basis. This
is a well established procedure and has at low energies (below the pion production
threshold) a clear physical meaning. At higher energies the number of partial waves
needed to obtain converged results increases, and approaches based on a direct
evaluation of the scattering equation in terms of vector variables become more
appealing. 

Especially the experience in three- and four-nucleon
calculations~\cite{wgrep,Nogga:2000uu} shows that the
standard treatment based on a partial wave projected momentum space basis is quite
successful at lower energies, but becomes increasingly more tedious with increasing
energy, since each building block requires extended algebra and intricate numerical
realizations. 
On the other hand for a system of three bosons interacting via scalar forces the
relative ease with which a three-body bound state~\cite{Elster:1998qv} as well as 
three-body scattering~\cite{Liu:2004tv}  can be calculated in the Faddeev scheme
when avoiding an angular momentum decomposition altogether has been successfully
demonstrated. Thus it is only natural to strive for solving the three nucleon (3N) Faddeev
equations in a similar fashion. 

Recently we proposed a three-dimensional (3D) 
formulation of the Faddeev equations
for 3N bound states~\cite{2N3N} and 3N scattering~\cite{3Nscatt} in which the 
spin-momentum operators are evaluated analytically, leaving the Faddeev equations as 
a finite set of coupled equations for scalar functions depending only on vector momenta.
One of the basic foundations of this formulation rests on the fact that the most
general form of the NN interaction
can only depend on six linearly independent spin-momentum operators, which in turn
dictate the form of the NN bound and scattering state. Here we extend the formulation
of the NN bound state given in~\cite{2N3N} to NN scattering and provide a numerical
realization.

There have been several approaches of formulating NN scattering without employing a
partial wave decomposition. A helicity formulation related to the total NN spin was
proposed in~\cite{imam}, which was extended to 3N bound state calculations
in~\cite{Bayegan:2007ih}. The spectator equation for relativistic NN scattering has
been successfully solved in~\cite{Ramalho:2006jk} using a helicity formulation. 
Aside from NN scattering, 3D formulations for the scattering of pions off
nucleons~\cite{Caia:2003ke} and protons off light nuclei~\cite{RodriguezGallardo:2007dc} 
have recently been successfully carried out.

In Section~\ref{section2} we introduce the formal structure of our approach starting
from the most general form of the NN potential. We derive the resulting Lippmann-Schwinger 
equation and show how to extract Wolfenstein parameters and NN scattering
observables.
 Numerical realizations of our approach that employ
a recent chiral next-to-next-leading order (NNLO) 
NN force~\cite{Ep05,evgeny.report,newer.report}
as well as the standard one-boson-exchange potential Bonn~B~\cite{machl} 
are presented in Section~\ref{section3}. 
The scalar functions, which result from the evaluation of the spin-momentum operators 
and have to be calculated only once are given in Appendices~\ref{appendixa} and
\ref{appendixb}. 
Finally we conclude in Section~\ref{conclusion}.
The more technical information necessary to perform calculations
with the chiral potential is given in Appendix~\ref{appendixc}.
In Appendix~\ref{appendixd} the Bonn~B potential is presented in the form 
required by our formulation.

\section{The Formal Structure}
\label{section2}

We start by projecting the NN potential on the NN isospin states 
$ \mid t m_t\rangle$, with $t=0, m_t=0$ being the singlet 
and $t=1, m_t = -1,0,1$ the triplet.
We assume that isospin is conserved, but allow for  charge independence and charge symmetry breaking,
and thus for a dependence on $m_t$,
\begin{eqnarray}
\langle  t' m_t' \mid V \mid t m_t\rangle  = 
  \delta_{ t t'} \delta_{ m_t m_t'} V^ { t m_t}
\label{eq:2.1}
\end{eqnarray}
Furthermore, the most general rotational, parity and time reversal invariant 
form of the off-shell NN force can be expanded into six scalar 
spin-momentum operators \cite{wolfenstein}, which we choose as
\begin{eqnarray}
w_1 ( {\bsigma}_1,{\bsigma}_2, {\bf p'}, {\bf p})&  = &  1\cr
w_2 ( {\bsigma}_1,{\bsigma}_2, {\bf p'}, {\bf p})&  = & {\bsigma}_1 \cdot {\bsigma}_2\cr
w_3 ( {\bsigma}_1,{\bsigma}_2, {\bf p'}, {\bf p)}&  = & i \; ( {\bsigma}_1
+ {\bsigma}_2 ) \cdot ( {\bf p} \times {\bf p'})\cr
w_4 ( {\bsigma}_1,{\bsigma}_2, {\bf p'}, {\bf p})&  = & {\bsigma}_1
\cdot ( {\bf p} \times {\bf p'}) \; {\bsigma}_2 \cdot ( {\bf p} \times {\bf p'})\cr
w_5 ( {\bsigma}_1,{\bsigma}_2, {\bf p'}, {\bf p})&  = & {\bsigma}_1
\cdot  ({\bf p'} + {\bf p}) \; {\bsigma}_2 \cdot  ({\bf p'} + {\bf p})\cr
w_6 ( {\bsigma}_1,{\bsigma}_2, {\bf p'}, {\bf p})&  = & {\bsigma}_1
\cdot ( {\bf p'} - {\bf p}) \; {\bsigma}_2 \cdot  ( {\bf p'} - {\bf p})
\label{eq:2.2}
\end{eqnarray}
Each of these operators is multiplied with scalar functions which depend only
on the momenta ${\bf p}$ and ${\bf p'}$, leading to the most general expansion for
any NN potential
\begin{eqnarray}
V^ { t m_t}  \equiv   \sum_{j=1}^ 6 v_j^ { t m_t} ({\bf p'}, {\bf p}) \;
w_j({\bsigma}_1,{\bsigma}_2, {\bf p'}, {\bf p})
\label{eq:2.3}
\end{eqnarray}
The property of Eq.~(\ref{eq:2.1}) carries over to the NN t-operator, which fulfills the
Lippmann-Schwinger (LS) equation
\begin{eqnarray}
t^ { t m_t} = V^ { t m_t} + V^ { t m_t} G_0 t^ { t m_t},
\label{eq:2.4}
\end{eqnarray}
with $G_0(z)=(z-H_0)^{-1}$ being the free resolvent. The t-matrix element has an expansion
analogous to the potential,
\begin{eqnarray}
t^ { t m_t}  \equiv   \sum_{j=1}^ 6 t_j^ { t m_t} ( {\bf p'}, {\bf p}) \;
 w_j( {\bsigma}_1,{\bsigma}_2, {\bf p'}, {\bf p})
\label{eq:2.5}
\end{eqnarray}
Inserting Eqs. (\ref{eq:2.3}) and (\ref{eq:2.5})  into the LS equation
(\ref{eq:2.4}), operating with  
$w_k( {\bsigma}_1,{\bsigma}_2, {\bf p'}, {\bf p}) $ from the left and performing the 
trace in the NN spin space leads to
\begin{eqnarray}
\lefteqn{ \sum_{j} A_{ kj} ( {\bf p'}, {\bf p}) t_j^{t m_t}({\bf p'}, {\bf p}) = 
 \sum_{j} A_{ kj} ( {\bf p'}, {\bf p}) v_j^ { t m_t}( {\bf p'}, {\bf p} )  } \cr
&+&  \int d^ 3 p'' \sum_{ j j'} v_j^ { t m_t}( {\bf p'}, {\bf p''}) G_0 ( p'') \;
t_{ j'}^ { t m_t}( {\bf p''}, {\bf p)} \; B_{ kjj'} ( {\bf p'}, {\bf p''},  {\bf p}).
\label{eq:2.6}
\end{eqnarray}
The scalar coefficients  $A_{kj}$ and $B_{kjj'}$ are defined as
\begin{eqnarray}
A_{ kj} ( {\bf p'}, {\bf p}) &\equiv &  {\rm Tr} \Big(w_k( {\bsigma}_1,{\bsigma}_2, {\bf
p'}, {\bf p}) \;  w_j({\bsigma}_1,{\bsigma}_2, {\bf p'}, {\bf p})\Big)
\label{eq:2.7} \\
 {B_{ k j j'} ( {\bf p'}, {\bf p''},  {\bf p}) } & \equiv  & 
  {\rm Tr} \Big(w_k( {\bsigma}_1,{\bsigma}_2, {\bf p'}, {\bf p}) \;
 w_j({\bsigma}_1,{\bsigma}_2, {\bf p'}, {\bf p''}) \;
  w_{j'}({\bsigma}_1,{\bsigma}_2, {\bf p''}, {\bf p})\Big)
\label{eq:2.8}
\end{eqnarray}
Here all spin dependencies are analytically evaluated, and the coefficients only depend on
the vectors ${\bf p}$, ${\bf p'}$, and ${\bf p''}$.  The explicit expressions for the
coefficients are given in Appendix~\ref{appendixa}. 

Thus we end up with a set of six coupled  equations for the scalar functions 
$t_j^ { t m_t}({\bf p'}, {\bf p})$, which depend for fixed $|{\bf p}|$ on two other variables, 
$ | {\bf p'}| $ and the cosine of the relative angle between the
vectors ${\bf p'}$ and ${\bf p}$, given by
${\bf \hat p'} \cdot {\bf \hat p}$.

Since Eqs.~(\ref{eq:2.3}) and (\ref{eq:2.5}) are completely general, 
any arbitrary NN force can be cast into this form and serve as input.
Finalizing the formulation, we only need to
antisymmetrized in the initial state by applying  
$( 1 - P_{12}) | {\bf p}\rangle | m_1 m_2\rangle | t m_t\rangle $, 
  and consider the on-shell t-matrix element for given $t m_t$:
\begin{eqnarray}
 M^ {t m_t}_{ m_1' m_2', m_1 m_2} &\equiv&
-\frac{m}2 (2 \pi)^2 \;
 t_{ m_1' m_2', m_1 m_2}^ { t m_t}(
{\bf p'}, {\bf p}) \Big|_{on-shell} \cr
& = & -\frac{m}2 (2 \pi)^2 \, \Big(   
\langle m_1' m_2'| \big[ t^ { t m_t}( {\bf p'}, {\bf p}) 
+ (-)^ t \; t^ { t m_t}( {\bf p'}, - {\bf p}) P_{12}^ s \big] | m_1 m_2\rangle 
\Big) .
\label{eq:2.9}
\end{eqnarray}
Here $P_{12}^ s$ interchanges the spin magnetic quantum numbers 
for the initial particles,  $m$ represents the nucleon mass.

For the on-shell condition, characterized by $ | {\bf p'}| = | {\bf p}|$ the vectors 
${\bf p}-{\bf p'}$ and ${\bf p}+{\bf p'}$ are orthogonal. Under this condition, the 
operator $ {\bsigma}_1 \cdot {\bsigma}_2$ can be represented as a linear combination 
of the operators $ w_j$, $j=4-6$~\cite{book}, i.e.
\begin{eqnarray}
{\bsigma}_1 \cdot {\bsigma}_2 & = &  \frac{ 1}{ ( {\bf p} \times {\bf p'})^ 2 }  \;
{\bsigma}_1 \cdot ( {\bf p} \times {\bf p'}) \; {\bsigma}_2 \cdot ( {\bf p}
\times {\bf p'})\cr
& + &  \frac{ 1}{ ( {\bf p} + {\bf p'})^ 2 } \; {\bsigma}_1 \cdot ( {\bf p} + {\bf
p'}) \; {\bsigma}_2 \cdot
( {\bf p} +  {\bf p'})\cr
& + &  \frac{ 1}{ ( {\bf p} - {\bf p'})^ 2 } \; {\bsigma}_1 \cdot ( {\bf p} - {\bf
p'}) \; {\bsigma}_2 \cdot ( {\bf p} -  {\bf p'})
\label{eq:2.10}
\end{eqnarray}
We can use the relation of Eq.~(\ref{eq:2.10})
for internal consistency checks of the calculations.
However, in order to keep the most general off-shell structure of Eq.~(\ref{eq:2.5}), we need
to keep all six terms. 
We will come back to the numerical implications of this fact below.

From Eq.~(\ref{eq:2.9}) we read off that the scattering matrix is given by
\begin{eqnarray}
 M^ {t m_t}_{ m_1' m_2', m_1 m_2} & = &  
-\frac{m}2 (2 \pi)^2 \,
\sum_{j=1}^ 6 \bigg[t_j^ { t m_t} ( {\bf p'},
{\bf p})  \;
\langle m_1' m_2'|  w_j( {\bsigma}_1,{\bsigma}_2, {\bf p'}, {\bf p})| m_1 m_2\rangle\cr
& + & (-)^ t t_j^ { t m_t} ( {\bf p'}, - {\bf p}) \;
 \langle m_1' m_2'|  w_j( {\bsigma}_1,{\bsigma}_2, {\bf p'},- {\bf p})  |m_2 m_1 \rangle \bigg]
\label{eq:2.11}
\end{eqnarray}
On the other hand the standard form  of the on-shell t-matrix 
for given quantum numbers $ t m_t$ \cite{book} reads in the Wolfenstein representation
\begin{eqnarray}
 M^ {t m_t}_{ m_1' m_2', m_1 m_2} &= & a^ {tm_t} \; \langle m_1' m_2'| m_1 m_2\rangle \cr 
&  - &  i \frac{ c^ {tm_t}}{ | {\bf p} \times {\bf p'}|} \;
 \langle m_1' m_2'|w_3 ({\bsigma}_1, {\bsigma}_2, {\bf p'}, {\bf p})| m_1 m_2\rangle\cr
&  + &  \frac{ m^ {tm_t}}{ | {\bf p} \times {\bf p'}|^ 2}
 \langle m_1' m_2'| w_4( {\bsigma}_1, {\bsigma}_2, {\bf p'}, {\bf p}) | m_1 m_2\rangle \cr
&  + &  \frac{ (g+h)^ {tm_t}}{ ( {\bf p} + {\bf p'})^ 2} \;
\langle m_1' m_2'|  w_5 ( {\bsigma}_1, {\bsigma}_2, {\bf p'},{\bf p}) | m_1 m_2\rangle \cr
&  + &  \frac{ (g-h)^ {tm_t}}{ ( {\bf p} - {\bf p'})^ 2}  \;
\langle m_1' m_2'| w_6( {\bsigma}_1, {\bsigma}_2, {\bf p'},{\bf p}) | m_1 m_2\rangle
\label{eq:2.12}
\end{eqnarray}
   
Due to  the action of $ P_{12}^s $ in Eq.~(\ref{eq:2.9}), which 
interchanges $m_1$ with $m_2$, 
the two parts of Eq.~(\ref{eq:2.11}) yield 
different results. 
Again, standard relations \cite{book,stapp} must be applied to extract
the Wolfenstein parameters:
\begin{eqnarray}
a^ {tm_t} & = &  \frac{1}{4} \; {\rm Tr} \left( M \right)\cr
c^ {tm_t} & = & -  i \frac{1}{8}\; {\rm Tr} \left( M \; \frac{w_3 ({\bsigma}_1, 
{\bsigma}_2, {\bf p}', {\bf p})}{ | {\bf p}
\times {\bf p'}|} \right)\cr
m^ {tm_t} & = &  \frac{1}{4} \; {\rm Tr} \left( M \; \frac{w_4 ({\bsigma}_1, {\bsigma}_2,
{\bf p'}, {\bf p})}{ | {\bf p}
\times {\bf p'}|^ 2} \right)\cr
(g+h)^ {tm_t} & = &  \frac{1}{4} \; {\rm Tr} \left( M \; \frac{w_5 ({\bsigma}_1, 
{\bsigma}_2, {\bf p}', {\bf p})}{ ( {\bf p}
+ {\bf p'})^ 2} \right)\cr
(g-h)^ {tm_t} & = &  \frac{1}{4} \; {\rm Tr} \left( M \; \frac{w_6 ({\bsigma}_1, 
{\bsigma}_2, {\bf p'}, {\bf p})}{ ( {\bf p}
- {\bf p'})^ 2} \right)
\label{eq:2.13}
\end{eqnarray}
It is straightforward to work out those relations starting from 
Eq.~(\ref{eq:2.11}).
In order to simplify the notation we
write $t_j \equiv t_j^ {t m_t} ( {\bf p'}, {\bf p})$, $\tilde t_j \equiv t_j^
{t m_t} ( {\bf p'}, -{\bf p})$, and  
$x \equiv {\bf \hat p'} \cdot {\bf \hat p}$  
and obtain 
\begin{eqnarray}
a^ {tm_t} & = &  t_1 + (-)^ t \; \Big[ \frac{1}{2} \tilde t_1 + \frac{3}{2} \tilde t_2 
 +   \frac{1}{2} p^ 4 ( 1-x^ 2)  \tilde t_4 + p^ 2 ( 1-x)  \tilde t_5 + p^ 2 (
1+x) \tilde t_6\Big] \cr
c^ {tm_t} & = &  i p^ 2 \sqrt{1 - x^ 2}  \; \left( t_3 - (-)^ t \tilde t_3 \right) \cr 
m^ {tm_t} & = &  t_2 + p^ 4 ( 1 - x^ 2)  t_4 
 +   (-)^ t \; \Big[ \frac{1}{2} \tilde t_1 - \frac{1}{2} \tilde t_2  + \frac{1}{2} p^ 4 ( 1- x^ 2) \tilde t_4 \cr
& & -   p^ 2 (1-x)\tilde t_5 - p^ 2 ( 1+x) \tilde t_6 \Big]\cr
g^ {tm_t} & = &  t_2 + p^ 2 ( 1+x)  t_5  + p^ 2( 1-x)  t_6 
 +  (-)^ t \; \Big[  \frac{1}{2} \tilde t_1 - \frac{1}{2} \tilde t_2  -
\frac{1}{2} p^ 4 ( 1-x^ 2) \tilde t_4 \Big]\cr
h^ {tm_t} & = &  p^ 2(1+x)  t_5 - p^ 2(1-x) t_6 
 +  (-)^ t \; \Big[- p^ 2(1-x)  \tilde t_5 + p^ 2(1+x)  \tilde t_6 \Big]
\label{eq:2.14}
\end{eqnarray}
It remains to consider the particle representation. For the proton-proton or 
neutron-neutron system the isospin is $t=1$. Thus the above given Wolfenstein 
parameters are already the physical ones and enter the calculation of
observables. In the case of the neutron-proton system both isospins contribute 
and the physical amplitudes are given by
  $ \frac{1}{2} ( a^ {00} + a^ {10})$, $\frac{1}{2} ( c^ {00} + c^ {10})$, etc.

Once the Wolfenstein parameters are known, all NN observables can readily be 
calculated taking well defined bilinear products thereof~\cite{book}. 
For example, the spin averaged differential cross section $I_0$ is given
as $\frac{1}{4} {\rm Tr} \; M M^\dagger $.
  
For completeness, we also give the derivation of the  
deuteron which carries isospin $t=0$ and total spin $s=1$. 
We employ the operator form from Ref.~\cite{imam},
\begin{eqnarray}
\langle {\bf p} | \Psi_{m_d}\rangle & = &  \left[ \phi_1(p) + \left({\bsigma}_1 \cdot {\bf
p} \;
{\bsigma}_2 \cdot {\bf p} - \frac{1}{3} p^2\right) \; \phi_2(p) \right] | 1 m_d\rangle  \cr
& \equiv &  \sum_{ k=1}^2 \phi_k(p) \;  b_k( {\bsigma}_1, {\bsigma}_2, {\bf
p}) | 1 m_d \rangle,
\label{dtwave}
\end{eqnarray}
where $| 1 m_d \rangle $ describes the state in which the two spin-$\frac{1}{2}$ states
are coupled to the total spin-1 and the  magnetic quantum number $m_d$. 
The definition of the operators $b_{k}$ can be easily read off 
the first line of Eq.~(\ref{dtwave}).
The two scalar functions $ \phi_1(p)$ and $ \phi_2(p)$ are related in a simple way
to the standard $s$- and $d$-wave components of the deuteron wave function,
$\psi_0 (p)$ and $\psi_2 (p)$  by \cite{imam}
\begin{eqnarray}
\psi_0 (p) &=& \phi_1 (p) , \cr
\psi_2 (p)& = &\frac{4 p^2} { 3 \sqrt{2}} \, \phi_2 (p)  .
\label{eq8.2}
\end{eqnarray}
Next we use the Schr\"odinger equation in integral form projected on isospin
states,
\begin{eqnarray}
\Psi_{m_d} = G_0 V^ {00} \Psi_{m_d}.
\end{eqnarray}
Inserting the explicit expression of Eq.~(\ref{dtwave}) we obtain
\begin{eqnarray}
\Big[\phi_1( p )&+& \left({\bsigma}_1 \cdot {\bf p} \; 
{\bsigma}_2 \cdot {\bf p}- \frac{1}{3} p^2 \right) \phi_2(p) \Big]
| 1 m_d \rangle = \cr
&&  \frac{1}{ E_d -\frac{p^2}{m}} \int d^3 p' \sum_{j=1}^6
v_j^ {00} ({\bf p},{\bf p'})\;  w_j({\bsigma}_1, {\bsigma}_2, {\bf p},{\bf
p'})\cr
 & \times &  \Big[\phi_1( p') + \left( {\bsigma}_1 \cdot {\bf p'} \; 
{\bsigma}_2 \cdot {\bf p'} - \frac{1}{3} {p'}^2\right) \phi_2(p') \Big]| 1 m_d
\rangle  ,
\label{eq:2:30}
\end{eqnarray}
where $E_d$ is the deuteron binding energy.
We remove the spin dependence by projecting from the left with
 $ \langle 1 m_d| b_k( {\bsigma}_1, {\bsigma}_2, {\bf p})$ and
 summing over $ m_d$. This leads to
\begin{eqnarray}
\lefteqn{\sum_{m_d =-1}^{1 }  \langle 1 m_d |  b_k( {\bsigma}_1, 
{\bsigma}_2, {\bf p})\sum_{ k'=1}^2 \phi_{k'}(p)  b_{k'}( {\bsigma}_1, {\bsigma}_2 ,
{\bf p})| 1 m_d \rangle =}  \cr
 &  & \frac{1}{ E_d -\frac{p^2}{m}} \sum_{ m_d =- 1}^{1 } \int d^3 p' \sum_{j=1}^6
v_j^{00}( {\bf p},{\bf p'}) w_j({\bsigma}_1, {\bsigma}_2, {\bf
p},{\bf p'}) \cr
& \times &  \sum_{ k''=1}^2 \phi_{k''}(p') \;   b_{k''}( {\bsigma}_1, {\bsigma}_2, {\bf p'})| 1 m_d \rangle  ~.
\label{eq:2.31}
\end{eqnarray}
Defining the scalar functions
\begin{eqnarray}
A^ d_{k k'} ( p) \equiv  \sum_{ m_d =- 1}^{1 } \langle 1 m_d| b_k
( {\bsigma}_1, {\bsigma}_2, {\bf p})  b_{k'}( {\bsigma}_1, 
{\bsigma}_2, {\bf p})| 1 m_d \rangle 
\end{eqnarray}
and
\begin{eqnarray}
B^ d_{ kjk''} ( {\bf p},{\bf p'}) \equiv 
 \sum_{ m_d =- 1}^{1 }  \langle 1
m_d| b_k( {\bsigma}_1, {\bsigma}_2, {\bf p}) w_j( {\bsigma}_1, 
{\bsigma}_2, {\bf p}, {\bf p'})  b_{k''}( {\bsigma}_1, 
{\bsigma}_2, {\bf p'})| 1 m_d \rangle,
\end{eqnarray}
we obtain for Eq.~(\ref{eq:2.31}) 
\begin{eqnarray}
\sum_{ k'=1}^2 A^ d_{k k'}(p) \phi_{k'} (p) =\frac{1}{ E_d
-\frac{p^2}{m}}\int d^3 p' \sum_{j=1}^6 v_j^ {00}( {\bf p},{\bf p'})\sum_{
k''=1}^2 B^ d_{ kjk''} ( {\bf p},{\bf p'})\phi_{k''}(p') ~.
\end{eqnarray}
Note that $A^ d_{k k'}$ and $B^ d_{ kjk''}$ are both independent 
of the interaction. Therefore, these coefficients can be prepared 
beforehand for all calculations of the deuteron bound state, 
which consists of two coupled equations for the 
functions $\phi_1(p)$ and $\phi_2(p)$.
The summation over $ m_d $ guarantees the scalar nature of the functions
$A^ d_{ k k'} ( p)$ and $B^d_{ kjk''} ( {\bf p},{\bf p'})$, 
which are given in  Appendix~\ref{appendixb}.
The azimuthal angle can be trivially integrated out, leading to the
final form of the deuteron equation
\begin{eqnarray}
\lefteqn{\sum_{ k'=1}^2 A^ d_{ k k'} ( p)\phi_{k'} (p) = } \cr
& &\frac{2 \pi}{ E_d -\frac{p^2}{m}}
\sum_{ k''=1}^2
\int_0^{\infty} dp' {p'}^2\phi_{k''}(p') \int_{-1}^{1} dx
\sum_{j=1}^6 v^{00}_j(p,p',x) B^ d_{ kjk''} ( p,p',x),
\label{eqdeut}
\end{eqnarray}
where 
$ x \equiv {\bf \hat p'} \cdot {\bf \hat p}$.

\section{Numerical realization}
\label{section3}

\subsection{The deuteron}

For a numerical treatment of Eq.~(\ref{eqdeut}), it is convenient 
to  first define
\begin{eqnarray}
Z_{k,k'} ( p,p' ) \equiv 
\int_{-1}^{1} dx
\sum_{j=1}^6 v^{00}_j(p,p',x) B^d_{ kjk'} ( p,p',x)
\end{eqnarray}
and then assume that the integral over $p'$ will be carried out with some 
choice of Gaussian points and weights 
$(p_j,g_j) \ {\rm with} \ j=1,2, \dots, N$.
This leads to
\begin{eqnarray}
\sum_{ k'=1}^2  \, \sum_{ j=1}^N \left( 
g_j p_j^2 Z_{k k'} (p_i,p_j) \, + \,\delta_{ij} \frac { p_j^2}{ 2m \pi} 
A^d_{kk'} (p_i) 
\right) \phi_{k'} (p_{j})  \ = \
E_d \sum_{ k'=1}^2  \frac1{2 \pi} A^d_{kk'} (p_i) \, \phi_{k'} (p_i) .
\label{eq:3.2}
\end{eqnarray}
Eq.~(\ref{eq:3.2}) can be written as a so-called generalized eigenvalue problem
\begin{eqnarray}
R \xi = E_d Y \xi ,
\end{eqnarray}
or
\begin{eqnarray}
\sum_{l'=1}^{2N} R_{l l'} \xi_{l'} = E_d  \sum_{l'=1}^{2N} Y_{l l'} \xi_{l'} ,
\end{eqnarray}
where 
\begin{eqnarray}
l &= & i+ (k-1)N , \nonumber \\
\xi_{l'} & = & \phi_{k'} (p_j) , \ \  l' = j+ (k'-1)N , \nonumber \\
R_{l l'} &= &
g_j p_j^2 Z_{k k'} (p_i,p_j) \, + \,\delta_{ij} \frac { p_j^2}{ 2m \pi}
A^d_{kk'} (p_i)  \nonumber \\
Y_{l l'} &= &\delta_{ij} \frac{1}{ 2\pi} A^d_{kk'} (p_i) .
\end{eqnarray}
Since $ A^d_{11} \ne 0$, $ A^d_{12} = A^d_{21} = 0$ and $ A^d_{22} \ne 0$,
the matrix $Y$ is diagonal and can be easily inverted,  we 
encounter an eigenvalue problem 
\begin{eqnarray}
\left( Y^{-1} R \right) \,  \xi = E_d \; \xi ,
\label{eq7}
\end{eqnarray}
which is of the same type and dimension as is being solved for the
deuteron wave function in a standard partial wave representation,
where one calculates  the $s$- and $d$-wave components, $\psi_0(p) $ and $\psi_2(p)$.
The connection between the two solutions, $( \phi_1 (p) , \phi_2(p) ) $ and $( \psi_0 (p) , \psi_2(p) ) $,
 given by Eqs.~(\ref{eq8.2})
provides a direct check of the numerical accuracy.

As a first example we use a chiral NNLO potential \cite{evgeny.report},
which for the convenience of the reader is briefly described in Appendix C.
For the specific calculation performed here  we take the neutron-proton version 
of this potential and employ the parameters 
listed in Table~\ref{tab:chiralpara}. 

\begin{table}
\caption{\label{tab:chiralpara} The parameters of the chiral potential 
    of Ref.~\cite{evgeny.report}
              in order NNLO. The LEC's are given for the cutoff combination 
              $\Lambda$= 600 MeV and $\tilde{\Lambda}$= 700 MeV. The pion decay 
              constant $F_{\pi}$ and masses are given in MeV. The constants $c_{i}$ 
are given in GeV$^{-1}$, $C_{S}$ and $C_{T}$ in GeV$^{-2}$ and the other
               $C_{i}$ in GeV$^{-4}$}

\begin{tabular}{|c|c|c|c|c|c|c|c|c|}
\hline
$g_A$ & $F_{\pi}$ &   $m_{\pi^0}$ &  $m_{\pi^\pm}$  &      $m$   & $c_1$  &  $c_3$
&  $c_4$ &   \cr 
1.29    &     92.4    &   134.977       &   139.570           & 938.919  & -0.81  &
-3.40  &   3.40  & \cr
\hline \hline 
$C_{S}$   &   $C_{T}$     &   $C_{1}$    &   $C_{2}$ &   $C_{3}$    & $C_{4}$       &  $C_{5}$ &   $C_{6}$    &   $C_{7}$      \cr 
-112.932  &    2.60161  &   385.633  & 1343.49  &    -121.543  &  -614.322   &  1269.04 &    -26.4880   &  -1385.12     \cr 
\hline
\end{tabular}
\end{table}

\begin{figure}[hp]\centering
\epsfig{file=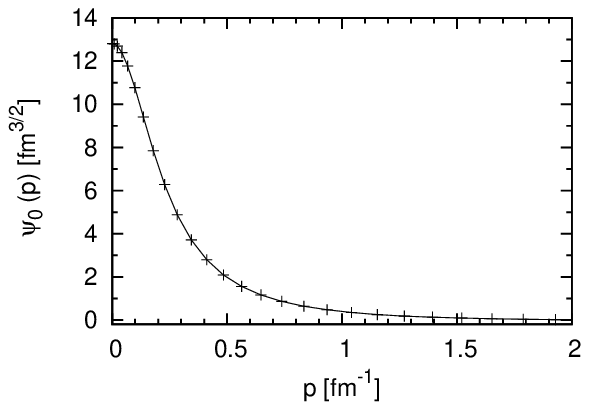,width=8cm,angle=0}
\epsfig{file=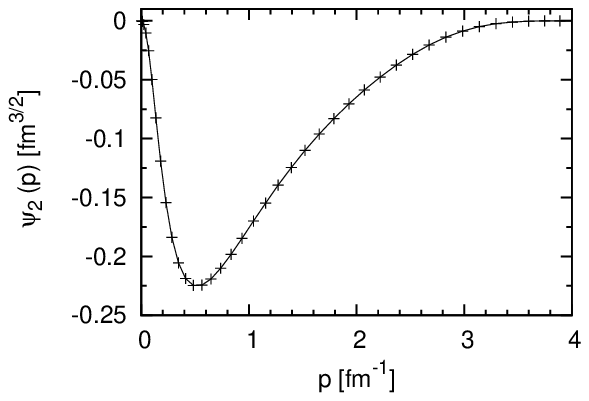,width=8cm,angle=0}
\caption{The $s$-wave (left) and $d$-wave (right) component of the deuteron
wave function as a function of the relative momentum $p$
for the chiral NNLO potential specified in the text. Crosses 
represent results obtained with the 
operator approach and solid lines are from the standard 
partial wave decomposition.}
\label{f1}
\end{figure}

We consistently use these potential parameters in 
the 3D and the PW calculations. In the first case we solve Eq.~(\ref{eq7}) for 
$\phi_1 (p) $ and $\phi_2 (p) $ and then use Eqs.~(\ref{eq8.2})
to obtain $\psi_0 (p) $ and $\psi_2 (p) $. In the second case we employ the 
standard partial wave representation of the potential and solve 
the Schr\"odinger equation directly for $\psi_0 (p) $ and $\psi_2 (p) $.
Both methods give the same value for the deuteron binding 
energy, namely $E_d$=-2.19993 MeV and  $s$-state probability $P_s$=95.291 \%. 
The wave functions are identical as can be seen 
in Fig.~\ref{f1}.

\begin{table}
\caption{\label{nntabobep}
Meson parameters for the Bonn B potential \cite{machl}. The $\sigma $ parameters shown in the table
are for NN total isospin 0. For NN total isospin 1 they should be replaced by $m_{\sigma } = 550$ MeV,
$\frac{g^2_{\alpha }}{4\pi } =  8.9437$, $\Lambda _{\alpha }= 1.9$ GeV and $n = 1$.}
{\par
\begin{center}
\begin{tabular}{|c||c|c|c|c|c|}
\hline
 meson & $m_{\alpha }$ [MeV] & $\frac{g^2_{\alpha }}{4\pi }$ & $\frac{f_{\alpha }}{g_{\alpha
}}$ & $\Lambda _{\alpha }$ [GeV] & n \\
\hline
\hline
 $\pi    $ & 138.03 & 14.4      &     & 1.7  & 1   \\
 $\eta   $ & 548.8  & 3         &     & 1.5  & 1   \\
 $\delta $ & 983    & 2.488     &     & 2    & 1   \\
 $\sigma $ & 720    & 18.3773   &     & 2    & 1   \\
 $\rho   $ & 769    & 0.9       & 6.1 & 1.85 & 2   \\
 $\omega $ & 782.6  & 24.5      & 0   & 1.85 & 2   \\
\hline
\end{tabular}
\end{center}
\par}
\label{tbonnb}
\end{table}

\begin{figure}[t]\centering
\epsfig{file=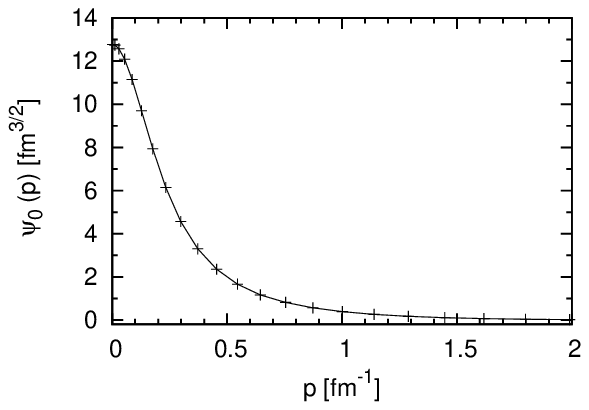,width=8cm,angle=0}
\epsfig{file=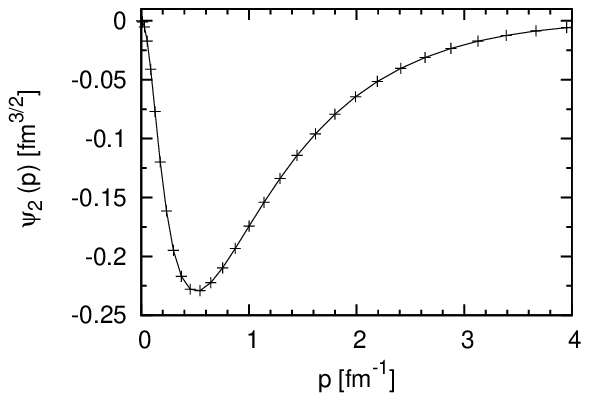,width=8cm,angle=0}
\caption{The same as in Fig.~\ref{f1} but for the Bonn B potential~\cite{machl}.}
\label{f2}
\end{figure}

As second NN force we choose the Bonn~B potential \cite{machl}, 
which has a more intricate structure due to the different
meson-exchanges and the Dirac spinors. The operator form of this potential,
corresponding to the basis of Eq.~(\ref{eq:2.2}) is derived 
in Appendix~\ref{appendixd} and the parameters 
are given in Table~\ref{tbonnb}.
In this case the nucleon mass is set to $m$= 939.039 MeV. 
Again we have an excellent agreement between the 3D and the partial wave based
calculation for the deuteron 
binding energy, $E_d$=-2.2242 MeV,  the $s$-state probability 
($P_s$= 95.014~\%) and the wave functions, which are displayed in Fig.~\ref{f2}.

In summary, we confirm that the 3D approach gives numerically 
stable results, which are in perfect agreement with the calculations based on 
standard partial wave methods.

\subsection{NN scattering observables}

The inhomogeneous LS equation (\ref{eq:2.6}) for the six 
components $ t_j^ {t m_t} $ can be solved for a fixed value of $p$.
For the vectors ${\bf {\hat p}}$ and ${\bf {\hat p}'}$ we choose
the explicit representation
\begin{eqnarray}
{\bf {\hat p}} &= &( 0,0,1 ) \nonumber \\
{\bf {\hat p}'} &= &( \sqrt{1 - {x'}^{\, 2}} ,0, x' ) \nonumber \\
{\bf {\hat p}''}& = &( \sqrt{1 - {x''}^{\, 2}} \cos
\varphi'',\sqrt{1 - {x''}^{\, 2}} \sin \varphi'',x'' )
\end{eqnarray}
so that the scalar products become
\begin{eqnarray}
{\bf {\hat p}'} \cdot {\bf {\hat p}}& =& x' \cr 
{\bf {\hat p}''}
\cdot {\bf {\hat p}} &= &x'' \cr 
{\bf {\hat p}'} \cdot {\bf {\hat
p}''} & = & x' x'' +  \sqrt{1 - {x'}^{\, 2}}  \sqrt{1 - {x''}^{\, 2}}
\cos \varphi'' \equiv y .
\end{eqnarray}

Let us now calculate the integral term on the right-hand-side of Eq.~(\ref{eq:2.6}) 
for a positive energy of the NN system, $E_{c.m.} \equiv \frac{p_0^2}{m}$:
\begin{eqnarray}
S_k (p',p,x') \equiv \int\limits_0^{\bar p}  dp'' {p''}^{\, 2}
\frac{1}{p_0^2 - {p''}^{\, 2} + i \epsilon } \; f_k ( p''; p',p,x') ,
\label{Sk}
\end{eqnarray}
where 
\begin{eqnarray}
 f_k ( p''; p',p,x') \equiv  f_k ( p'' ) \equiv  \nonumber \\
 m \sum\limits_{ j, j'=1}^6 \,
\int\limits_{-1}^1 d x'' \int\limits_0^{2 \pi} d\varphi'' \;
 B_{kjj'} ( p', p'', p, x',x'',\varphi'') \; v_j( p', p'',y) \;
t_{ j'}( p'', p, x'' ) \ .
\end{eqnarray}
Here the index $t m_t$ for the t-matrix element is omitted for simplicity.
For the momentum integration in Eq.~(\ref{Sk}) an
upper bound $\bar p$ is introduced, since the
 contributions to the integral 
for larger momenta are insignificant, the potential and 
the $t$-matrix  are essentially zero. Then  
the integral of Eq.~(\ref{Sk}) can be treated in a standard fashion and one
obtains
\begin{eqnarray}
S_k (p',p,x') = \int\limits_0^{\bar p}  dp'' \, \frac{ {p''}^{\,
2} f_k (p'' ) - p_0^2 f_k (p_0) }{p_0^2 - {p''}^{\, 2} } \, + \,
\frac12 p_0 f_k (p_0) \, \left( \ln  \frac{{\bar p}+p_0}{{\bar
p}-p_0} -i \pi \right) .
 \label{Sk2}
\end{eqnarray}
It is tempting to solve Eq.~(\ref{eq:2.6}) by iteration and 
then sum the resulting Neumann series with a Pad\'e scheme.
The determinant of the $6 \times 6$ matrix $A(p',p,x')$,
which appears on both sides of (\ref{eq:2.6}),  
can be easily calculated with the result
\begin{eqnarray}
\det (A) = -65536 \, p^8 \, {p'}^{\, 8} 
\left( p^2 -  {p'}^{\, 2} \right)^2 
\left( 1 -  {x'}^{\, 2} \right)^4 .
\label{detA}
\end{eqnarray}
In particular, this determinant is zero for $p' = p$ and $x' = \pm 1$.
However, by a careful choice of the $p$, $p'$ and $x'$ points,
it is possible to work with non-zero values of $\det(A)$, 
so that the matrix $A$ can be inverted. In this case Eq.~(\ref{eq:2.6})
can be written as 
\begin{eqnarray}
t(p',p,x') = v(p',p,x') + A^{-1}(p',p,x') \;  S(p',p,x') ,
\label{it1}
\end{eqnarray}
where $t(p',p,x')$, $v(p',p,x')$ and $S(p',p,x')$ denote now six-dimensional 
vectors with components $t_j$,  $v_j$ and $S_j$.
Note that $S(p',p,x')$ contains the unknown vector $t(p',p,x')$.
We arrive at the following iteration scheme:
\begin{eqnarray}
 t^{(1)} (p',p,x')& =& v(p',p,x') \nonumber \\
 t^{(n)} (p',p,x')& =& v(p',p,x') +  A^{-1}(p',p,x') S^{(n-1)} (p',p,x') ,
\ \ \ {\rm for} \ n > 1 ,
\label{it2}
\end{eqnarray}
where $S^{(n-1)} (p',p,x')$ is calculated using the vector $t(p',p,x')$ 
from the previous iteration, i.e. $t^{(n-1)} (p',p,x') $.
However, our experience with this iteration scheme is discouraging.
Numerically $\det(A)$ can be very close to zero, and in such cases 
the rank of matrix $A$ can vary from $2$ to $5$. 
As a consequence, it is 
very difficult to maintain numerical 
stability for this iterative method. Another drawback of using 
the inverse of $A$ is that it is impossible to obtain the on-shell 
matrix element $t(p_0,p_0,x')$ directly. One would have to rely on
numerical interpolations for calculating on-shell matrix elements.

For this reason we decided to solve Eq.~(\ref{eq:2.6})
directly as a system of inhomogeneous coupled algebraic equations.
To this aim we first perform a discretization  with respect to the
different variables in the problem.
As typical grid sizes 
we take $n_x = 36 $ Gaussian points for the $x''$ integration, and
 use the same grid for the $x'$ points.
Furthermore, we use $n_p = 36$  Gaussian points 
for the $p'$ and $p''$ grids, which are defined  the
interval $(0, {\bar p}= 40\, {\rm fm}^{-1})$. These points are distributed 
in such a way that $p_0$ is avoided and the same number of points is 
put symmetrically into two narrow intervals on each side of
$p_0$~\cite{ISloan}. 
Such a choice proved advantageous in the treatment of the 
$^1$S$_{0}$ channel for the PWD calculations and is kept here. 
In addition, $p_0$ is added to the set of $p'$ points.
Finally, we choose $n_{\varphi''}=60$ Gaussian
points for the $\varphi''$ integration. 
Thus, we arrive at a system of $6 \times (n_p+1) \times n_x$  linear equations
of the form
\begin{eqnarray}
H \xi = b,
\label{Hxb} 
\end{eqnarray}
where the vector $\xi$ represents all unknown values of 
$ t_j (p',p,x')$ for fixed $p$.
If we choose from the very beginning $p=p_0$, then 
the solution of Eq.~(\ref{Hxb}) 
contains the on-shell t-matrix in the operator 
form, namely $t_j(p_0,p_0,x')$. 

It is clear that for the on-shell $t$-matrix the solution cannot 
be unique, since the six operators become linearly 
dependent on each other (see Eq.~(\ref{eq:2.10})). 
In principle, one therefore expects that Eq.~(\ref{Hxb})  
is non-invertible and that tools like a singular value 
decomposition are required for the solution. However, we found 
that this is not required since the standard LU decomposition 
of {\it Numerical Recipes} \cite{numrec} worked safely for 
both interactions, all the considered laboratory energies 
and different choices of the 
mesh points. Interestingly, 
the actual solution for the on-shell $t$-matrix is not unique 
as expected and depends even on the optimization level 
of the compiler. However, the observables turn out to be stable and unique. 

Of course, setting $p= p_0$ is not necessary.
For $p \ne p_0$ the system of equations (\ref{Hxb}) has a unique 
and smooth solution and afterwards the interpolation to the on shell case 
can be safely performed.

The path to NN observables is straightforward.
From Eq.~(\ref{eq:2.11}) we evaluate first the scattering matrix $M$
for all possible spin projections $ m_1'$, $ m_2'$, $m_1$, and $m_2$,
noting that on-shell
\begin{eqnarray}
t_j^ { t m_t} ( {\bf p'},  {\bf p}) = t_j^ { t m_t} ( p_0, p_0, x')
\label{eqM2}
\end{eqnarray}
and
\begin{eqnarray}
t_j^ { t m_t} ( {\bf p'}, - {\bf p}) = t_j^ { t m_t} ( p_0, p_0, -x' ) .
 \label{eqM3}
\end{eqnarray}
Since we use a set of $x'$ points which is symmetric with respect
to $x'=0$, no interpolation is required and $M$ is easily
obtained.
Before we can make use of Eq.~(\ref{eq:2.13}),
we calculate matrix elements
of the modified operators $w_j$ appearing in (\ref{eq:2.13}), 
in the same representation as for the matrix $M$:
\begin{eqnarray}
\Big\langle m_1' m_2' \Big| \frac{w_3 ({\bsigma}_1, {\bsigma}_2, {\bf p}', {\bf p})}
{ | {\bf p} \times {\bf p'}|} \Big| m_1 m_2 \Big\rangle  \nonumber \\
\Big\langle m_1' m_2' \Big| \frac{w_4 ({\bsigma}_1, {\bsigma}_2, {\bf
p'}, {\bf p})}
{ | {\bf p} \times {\bf p'}|^ 2}  \Big| m_1 m_2 \Big\rangle \nonumber \\
\Big\langle m_1' m_2' \Big| \frac{w_5 ({\bsigma}_1, {\bsigma}_2, {\bf
p}', {\bf p})}
{ ( {\bf p} + {\bf p'})^ 2}  \Big| m_1 m_2 \Big\rangle \nonumber \\
\Big\langle m_1' m_2' \Big| \frac{w_6 ({\bsigma}_1, {\bsigma}_2, {\bf
p'}, {\bf p})} { ( {\bf p} - {\bf p'})^ 2}  \Big| m_1 m_2 \Big\rangle .
\label{eq:wlf1}
\end{eqnarray}
For this calculation symbolic software like
{\em Mathematica$^\copyright$}~\cite{math} proves very useful.
In the next step, the Wolfenstein parameters are calculated as sums over $ m_1'$, $
m_2'$, $m_1$ and $m_2$. For example
\begin{eqnarray}
& & a^ {tm_t} =  \frac{1}{4} \sum\limits_{m_1',m_2'} \sum\limits_{m_1
,m_2}
  M^{t m_t}_{ m_1' m_2' ,  m_1 m_2 } \, \delta_{m_1' \, m_1} \,
  \delta_{m_2' \, m_2} , \nonumber \\
& &   c^ {tm_t}  =  -  i \frac{1}{8}
  \sum\limits_{m_1',m_2'}
\sum\limits_{m_1 ,m_2}
  M^{t m_t}_{ m_1' m_2' ,  m_1 m_2 } \,
\Big\langle m_1 m_2 \Big| \frac{w_3 ({\bsigma}_1, {\bsigma}_2, {\bf p}',
{\bf p})} { | {\bf p} \times {\bf p'}|} \Big| m_1' m_2' \Big\rangle  .
\end{eqnarray}
Finally, the NN observables 
result from the Wolfenstein parameters 
as simple bilinear expressions \cite{book}.

\begin{figure}[t]\centering
\epsfig{file=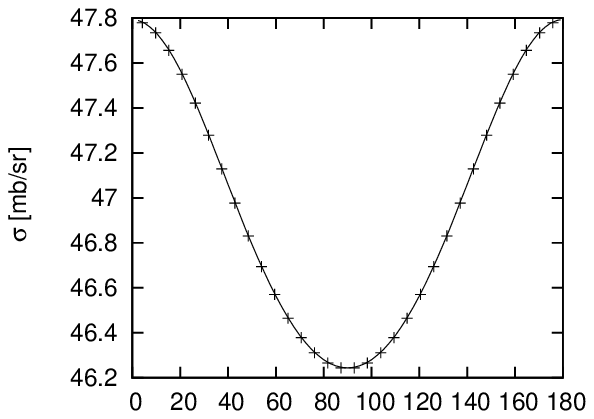,width=7cm,angle=0}
\epsfig{file=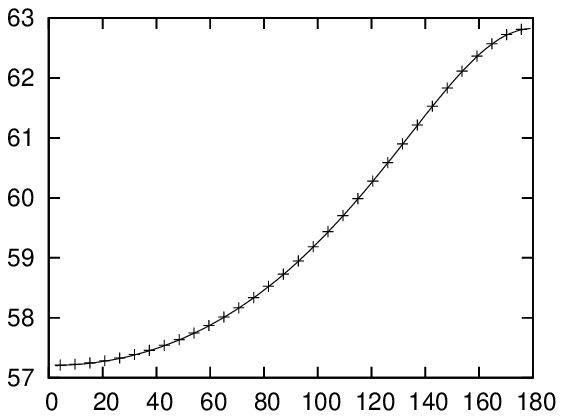,width=7cm,angle=0}
\epsfig{file=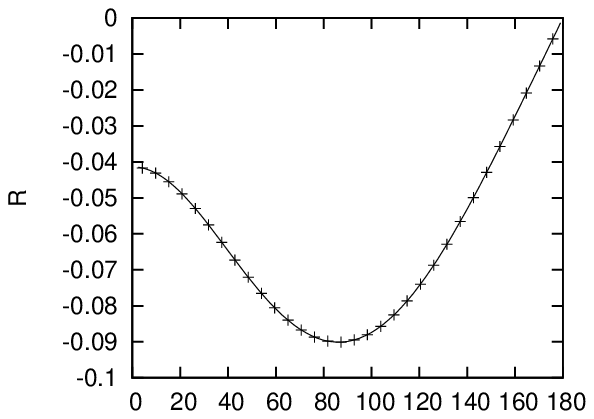,width=7cm,angle=0}
\epsfig{file=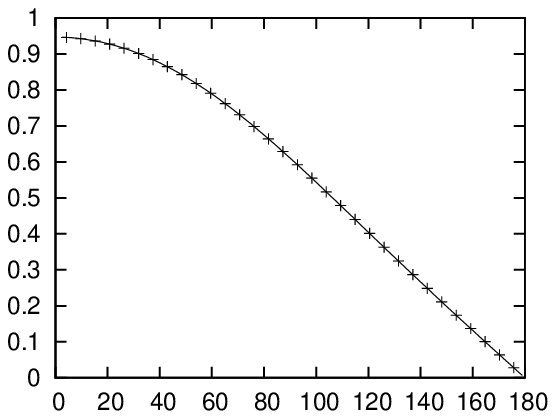,width=7cm,angle=0}
\epsfig{file=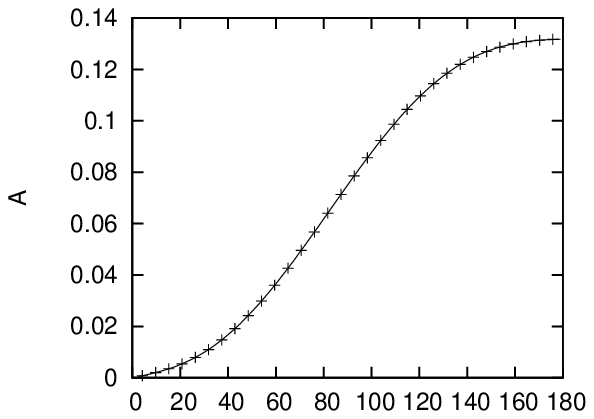,width=7cm,angle=0}
\epsfig{file=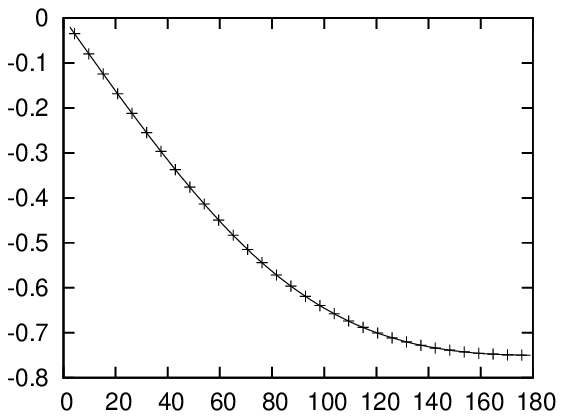,width=7cm,angle=0}
\epsfig{file=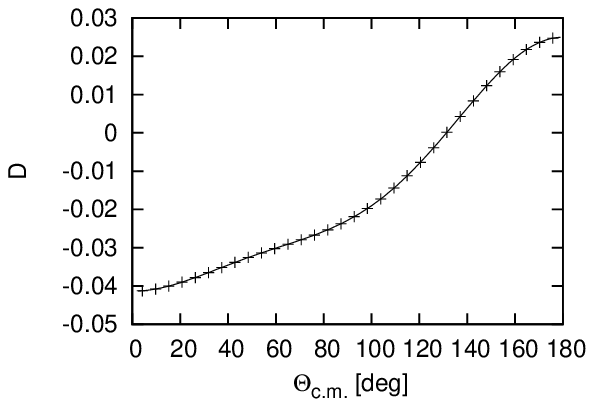,width=7cm,angle=0}
\epsfig{file=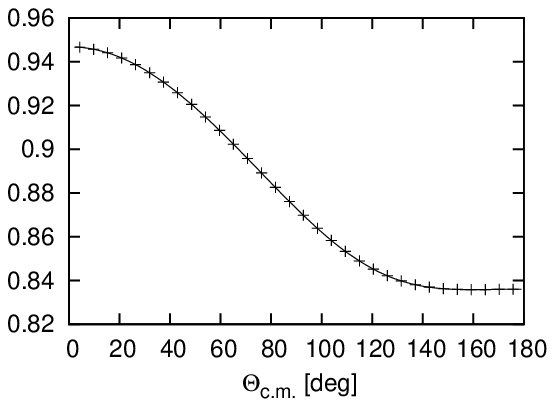,width=7cm,angle=0}
\caption{Selected observables for the neutron-neutron (left panel) 
and neutron-proton (right panel) system at the projectile laboratory kinetic energy 13~MeV
as a function of the center of mass angle $\theta$
for the  chiral NNLO potential~\cite{evgeny.report}. Crosses 
represent results obtained with the 
operator approach and solid lines represent fully converged results 
from the standard PWD. For the definition of the $R$, $A$ and $D$ observables 
see e.g.~\cite{book}.}
\label{f3}
\end{figure}

\begin{figure}[t]\centering
\epsfig{file=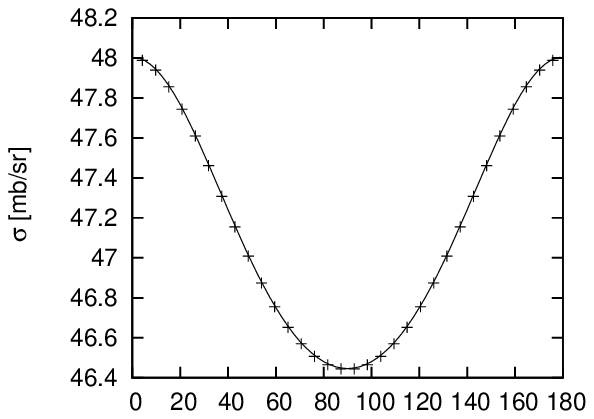,width=7cm,angle=0}
\epsfig{file=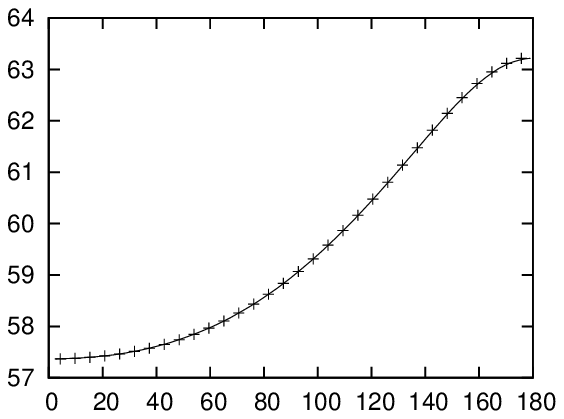,width=7cm,angle=0}
\epsfig{file=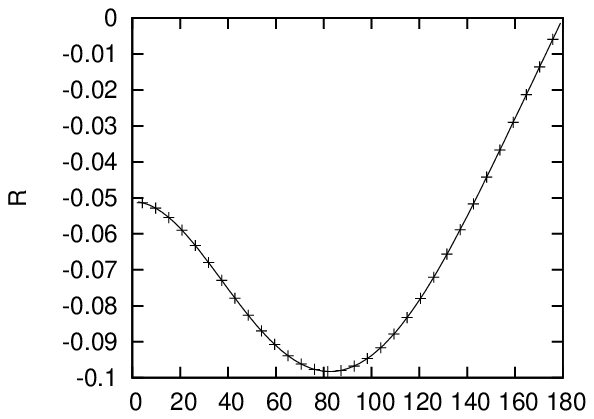,width=7cm,angle=0}
\epsfig{file=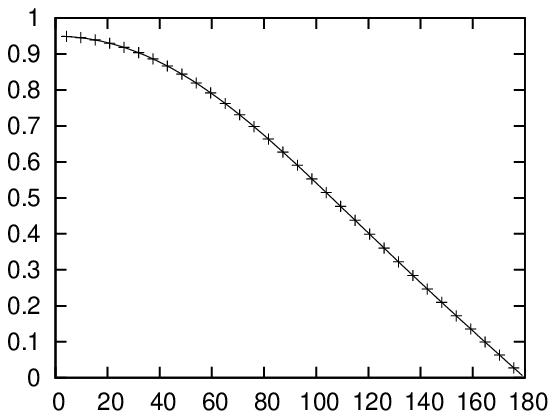,width=7cm,angle=0}
\epsfig{file=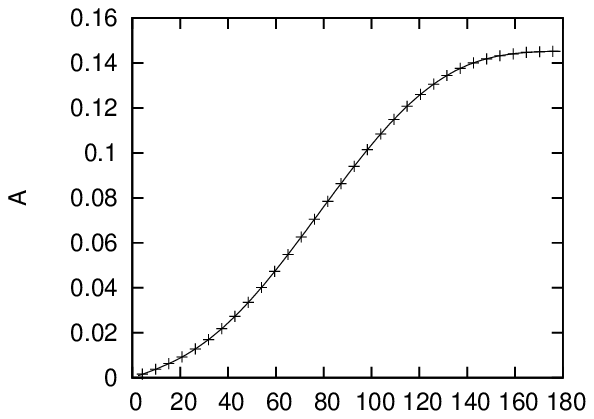,width=7cm,angle=0}
\epsfig{file=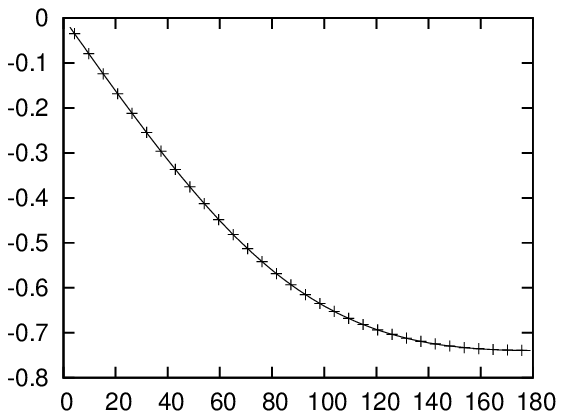,width=7cm,angle=0}
\epsfig{file=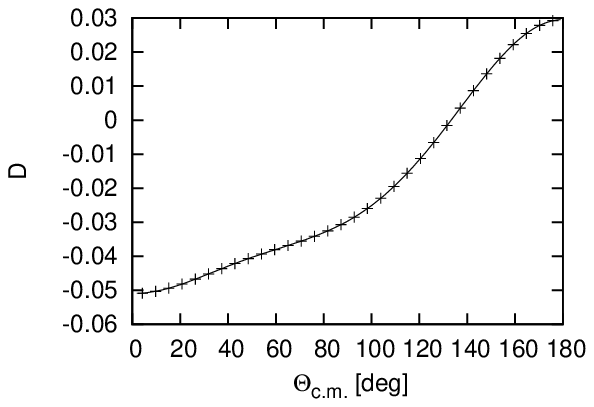,width=7cm,angle=0}
\epsfig{file=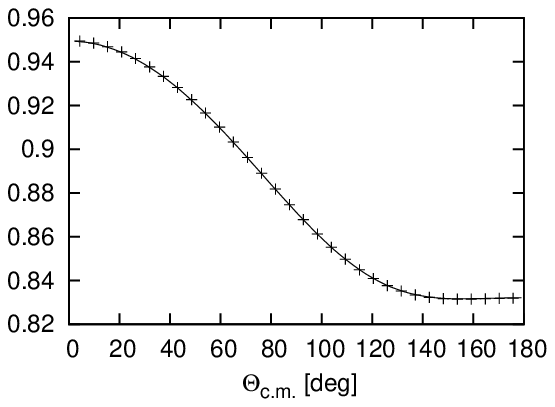,width=7cm,angle=0}
\caption{The same as in Fig.~\ref{f3} for the Bonn~B potential~\cite{machl}.}
\label{f4}
\end{figure}

\begin{figure}[t]\centering
\epsfig{file=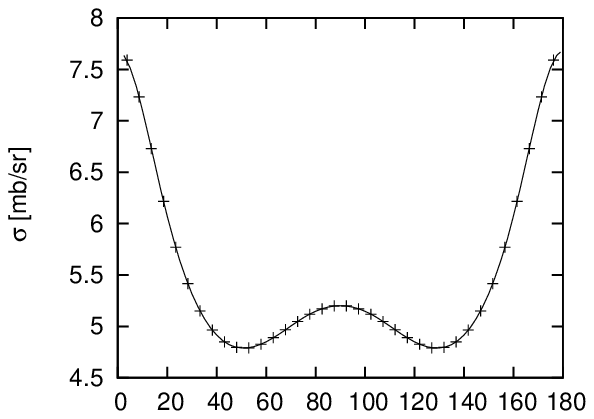,width=7cm,angle=0}
\epsfig{file=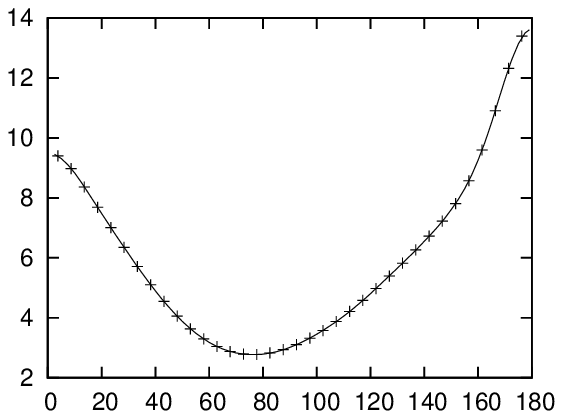,width=7cm,angle=0}
\epsfig{file=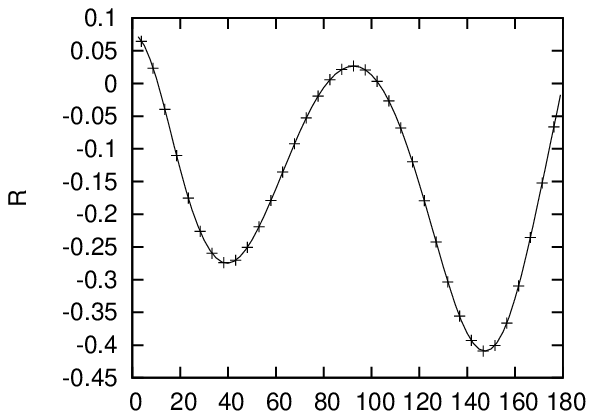,width=7cm,angle=0}
\epsfig{file=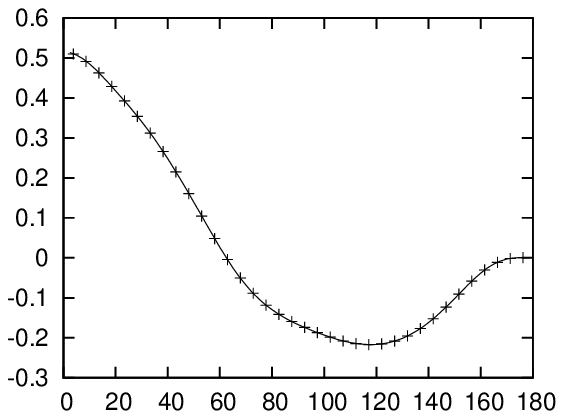,width=7cm,angle=0}
\epsfig{file=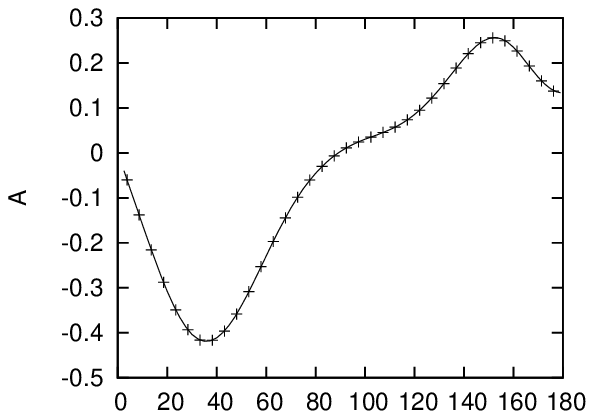,width=7cm,angle=0}
\epsfig{file=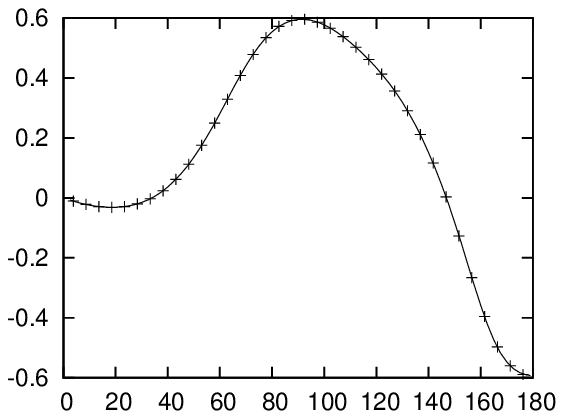,width=7cm,angle=0}
\epsfig{file=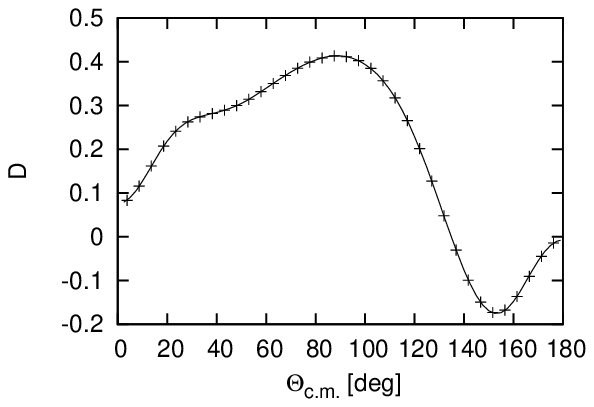,width=7cm,angle=0}
\epsfig{file=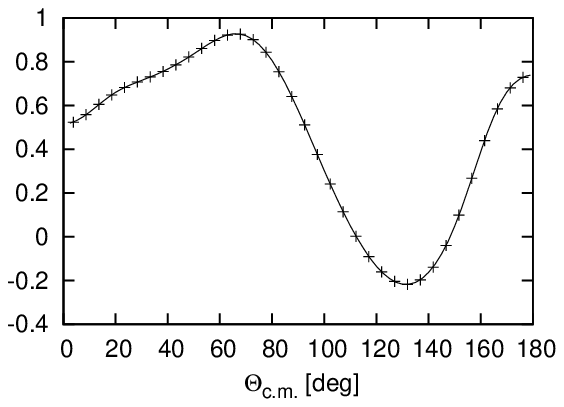,width=7cm,angle=0}
\caption{The same as in Fig.~\ref{f3} for the projectile laboratory kinetic energy
being  150~MeV.}
\label{f5}
\end{figure}

\begin{figure}[t]\centering
\epsfig{file=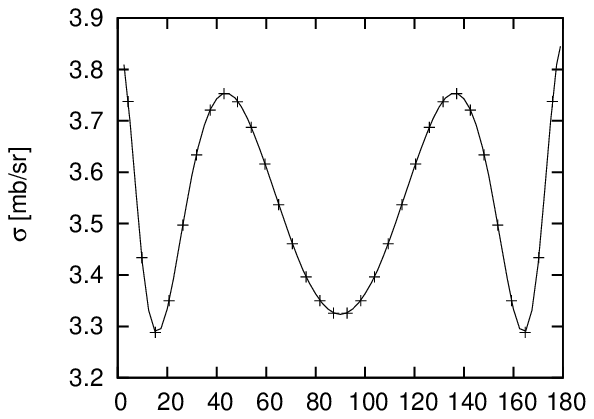,width=7cm,angle=0}
\epsfig{file=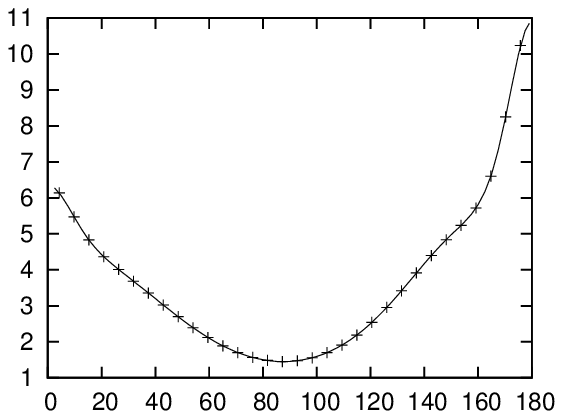,width=7cm,angle=0}
\epsfig{file=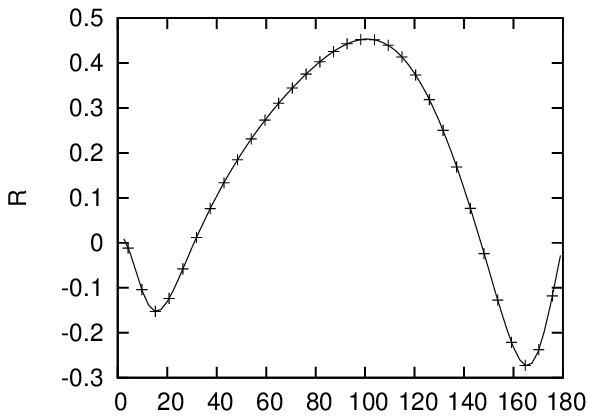,width=7cm,angle=0}
\epsfig{file=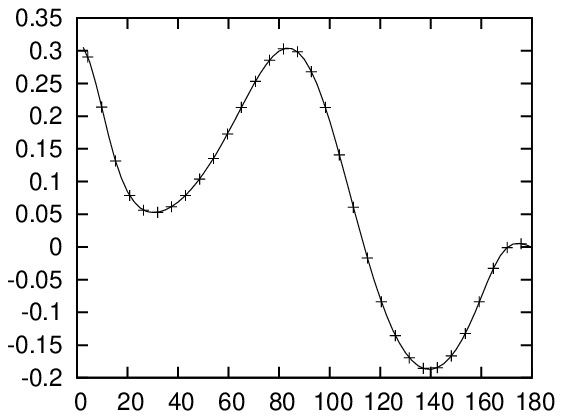,width=7cm,angle=0}
\epsfig{file=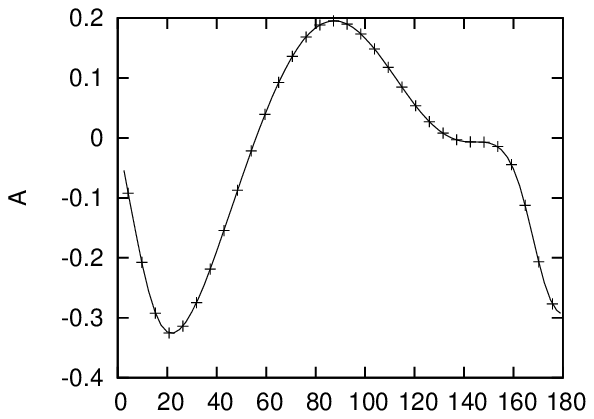,width=7cm,angle=0}
\epsfig{file=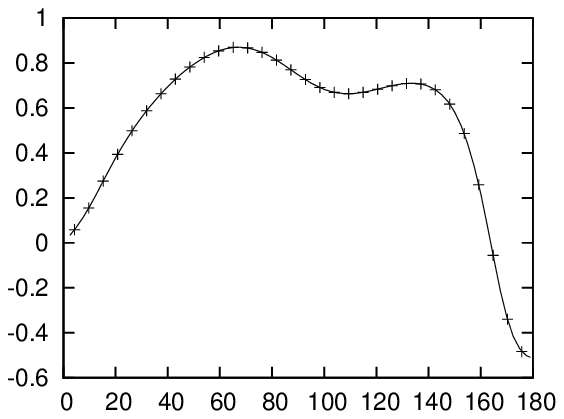,width=7cm,angle=0}
\epsfig{file=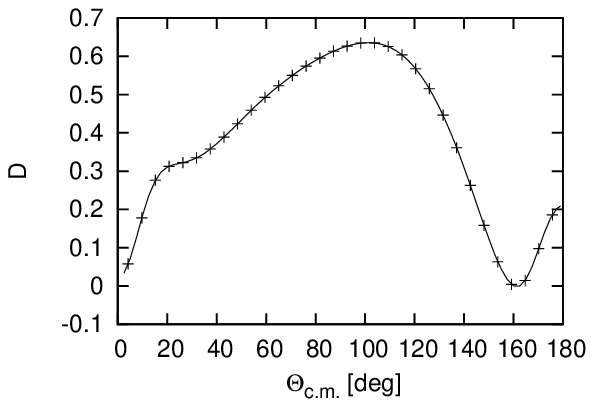,width=7cm,angle=0}
\epsfig{file=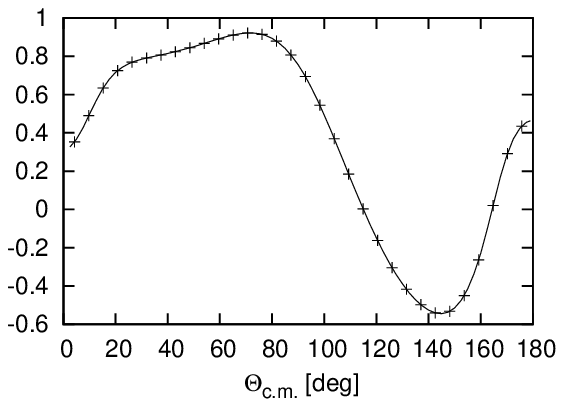,width=7cm,angle=0}
\caption{The same as in Fig.~\ref{f4} for the projectile laboratory kinetic energy 
being 300~MeV.}
\label{f6}
\end{figure}

In Figs.~\ref{f3}--\ref{f6} we compare a selected set of observables calculated 
with the new 3D method  to results obtained  by using a standard 
partial wave decomposition, employing the same potentials we used for the deuteron
calculations.
For the chiral potential we chose two laboratory kinetic energies 13 and 150~MeV,
whereas for the Bonn~B potential the higher energy is chosen to be 300~MeV.
We made sure that in all cases a sufficient number of partial waves is included 
to obtain converged results in the standard PWD approach. 
For all the  energies considered our converged PWD results agree perfectly
with predictions obtained from the new 3D approach.

\begin{figure}[t]\centering
\epsfig{file=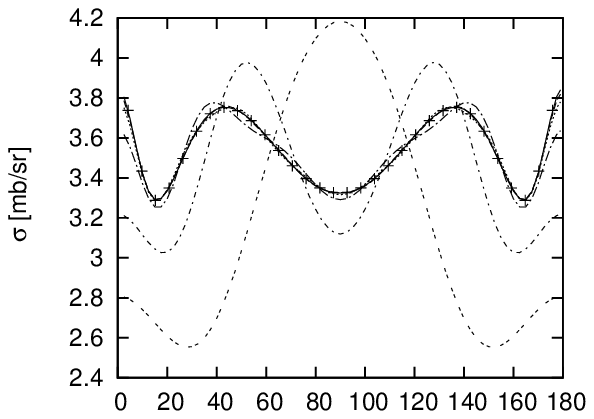,width=12cm,angle=0}
\epsfig{file=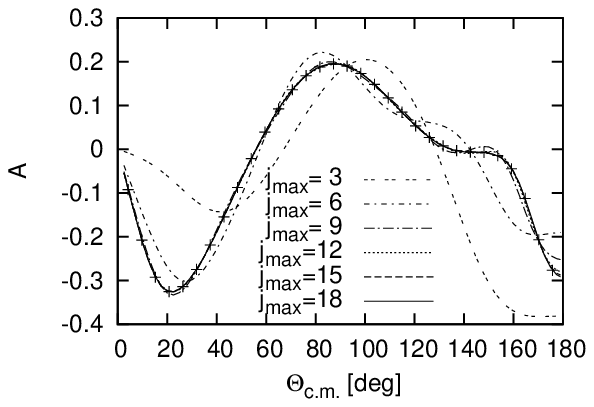,width=12cm,angle=0}
\caption{The convergence of the PWD results for the differential 
cross section and the depolarization coefficient $A$~\cite{book} for
neutron-neutron scattering
based on different numbers of 
partial waves determined by the maximal total angular momentum $j_{max}$ 
of the NN system (lines) with respect to the 
result of the three-dimensional calculation (crosses) for projectile laboratory 
kinetic energy 300~MeV and the Bonn~B potential~\cite{machl}.} 
\label{f7}
\end{figure}

\begin{figure}[t]\centering
\epsfig{file=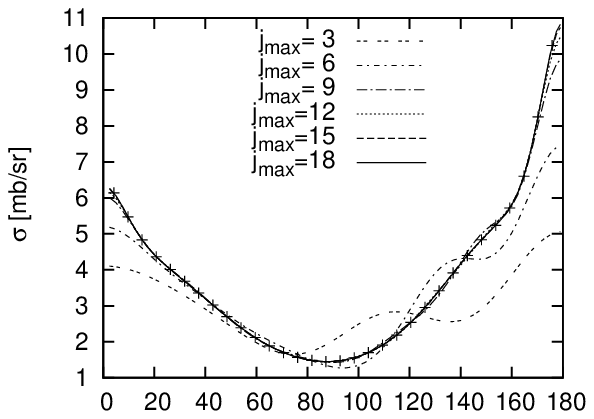,width=12cm,angle=0}
\epsfig{file=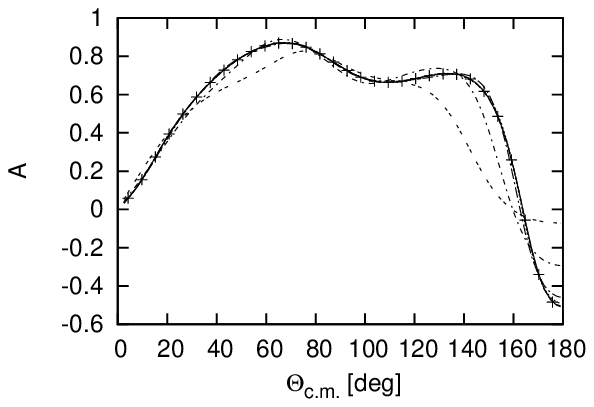,width=12cm,angle=0}
\caption{The same as in Fig.~\ref{f7} for neutron-proton scattering.}
\label{f8}
\end{figure}

\begin{figure}[t]\centering
\epsfig{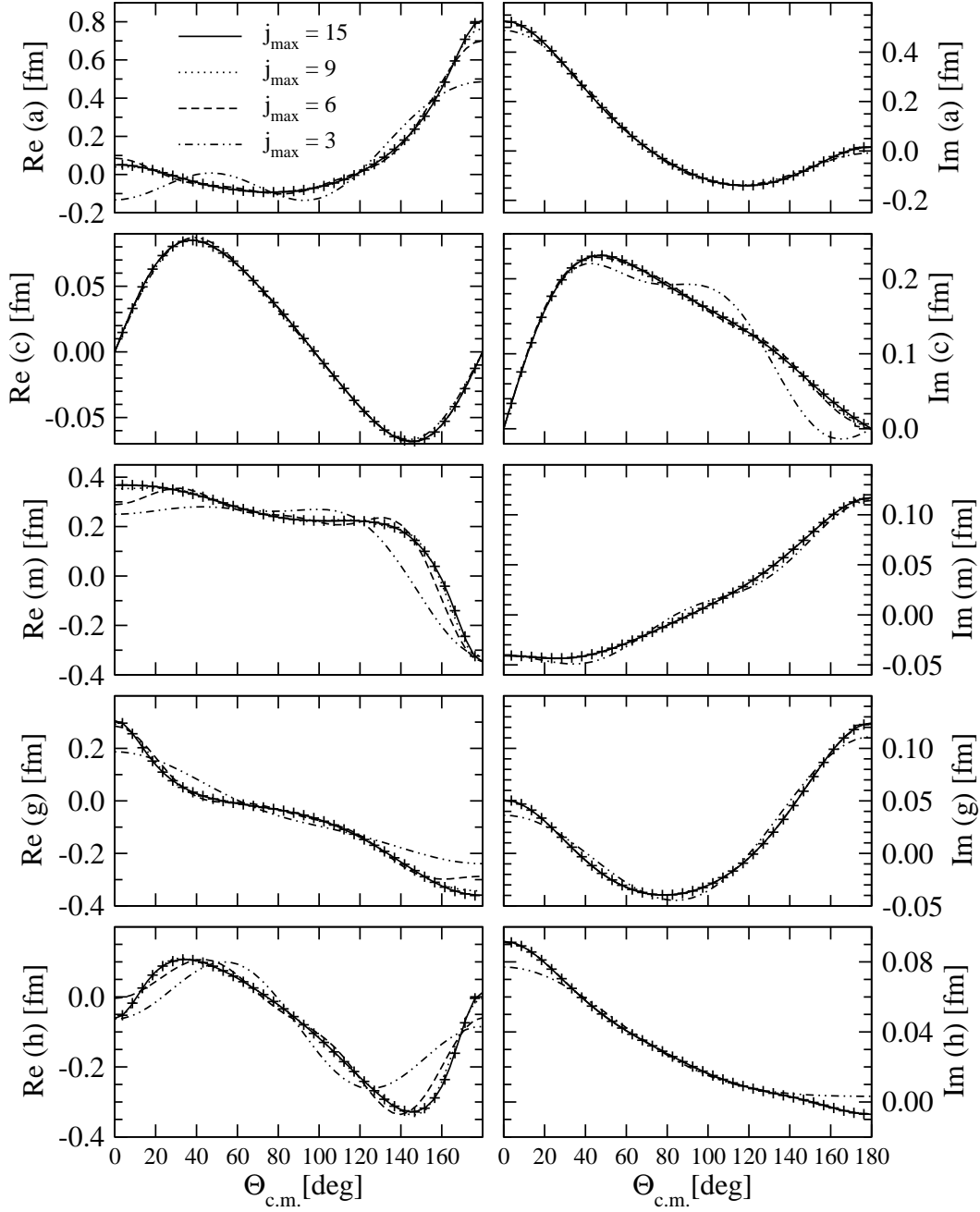}
\caption{The Wolfenstein parameters for neutron-proton scattering for projectile
laboratory kinetic energy 300~MeV calculated with the Bonn~B potential~\cite{machl}.
Results of the 3D calculation are given by the crosses. The convergence of the PWD results for
increasing values of maximum angular momentum $j_{max}$ is shown by the different curves
labeled in the figure. The left panels show the real parts of the amplitudes, whereas the
imaginary parts are displayed in the right panels.}
\label{f9}
\end{figure}

In Figs.~\ref{f7}--\ref{f8} we demonstrate the convergence with respect to
 different maximum total angular momenta $j_{max}$ 
towards the results calculated using our new 3D method 
for the differential cross section 
and the asymmetry $A$. Here we employ the Bonn~B potential and show the 
calculations for  
the neutron-neutron and neutron-proton cases separately.
As one can see,  quite a 
sizeable number of partial waves is required for a converged calculation
at 300~MeV. Finally, in Fig.~\ref{f9} we display the Wolfenstein amplitudes for
neutron-proton scattering at 300~MeV laboratory kinetic energy. Again we compare
partial wave based calculations for different maximum total angular momenta $j_{max}$ to
the 3D calculation. We observe that the maximum number of partial waves needed for
obtaining a converged result is quite different for the different amplitudes. 

\section{Summary and conclusions}
\label{conclusion}

Two nucleon scattering at intermediate energies of a few hundred MeV
requires quite a few angular momentum states in order to achieve convergence
of e.g. scattering observables. We formulated and numerically illustrated an
approach to treating the NN system working directly with momentum vectors  and using
spin-momentum operators multiplied by scalar functions, which only depend on the
momentum vectors. This approach is quite natural, since any general NN force
being invariant under time-reversal, parity and Galilei (or Lorentz) transformations
can only depend on six linear independent spin-momentum operators. 
The representation of the NN potential using spin-momentum operators leads to 
a system of six coupled equations of scalar functions (depending on momentum vectors)
for the NN t-matrix, once the spin-momentum operators are analytically calculated by
performing suitable trace operations. 

We calculated deuteron properties and NN scattering observables using two
different NN potentials, one derived from chiral effective field theory and one
from meson exchange. For all cases we found perfect agreement between the
calculations based on our new method and conventional calculations using a partial wave
basis.

This work is intended to serve as starting point towards treating three-nucleon systems
without partial waves. The theoretical formulation has already been given for the 3N
bound state (including 3N forces) in \cite{2N3N} and for 3N scattering in
\cite{3Nscatt}. For a much simpler  case when spin- and isospin-degrees are neglected the
feasibility of three-body scattering calculation in the GeV regime has already been
demonstrated~\cite{Liu:2004tv}, even including Poincar\'e symmetry~\cite{tinglin}.
Since our approach leads to coupled equations of scalar 
functions of momentum vectors, the
generalization to include spin-degrees of freedom appears feasible. 
In the  3N system not only the number of partial waves increases rapidly, 
but also 3N forces appear as new dynamical input. In particular
in the chiral approach the number of 3NF contributions proliferates
with order of expansion of the theory. In higher orders many complicated terms contribute 
to the 3N force \cite{evgeny.report}. 
In this case a traditional partial  wave decomposition of the 3NF poses a serious problem which 
has a chance to be alleviated by a 
direct three-dimensional treatment.


\section*{Acknowledgments}
We thank Dr. Evgeny Epelbaum for providing us with a code for the operator 
form of the chiral NNLO potential.

This work was supported by 
the Polish Ministry of Science and Higher Education
under Grants No. N N202 104536 and No. N N202 077435
and in part under the
auspices of the U.~S.  Department of Energy, Office of
Nuclear Physics under contract No. DE-FG02-93ER40756
with Ohio University.
It was also partially supported by the Helmholtz
Association through funds provided to the virtual institute ``Spin
and strong QCD''(VH-VI-231).
The numerical calculations were partly performed on the supercomputer 
cluster of the JSC, J\"ulich, Germany.


\appendix

\section{Coefficients for NN scattering}
\label{appendixa}

In this appendix we present the expressions $A$ and $B$ given in Eqs.~(\ref{eq:2.7}) 
and (\ref{eq:2.8}) for NN scattering.
The coefficients $A_{ij}({\bf p}', {\bf p})$ can be obtained in terms of the 
following four functions $FA$
\begin{eqnarray}
FA_{3}({\bf p}', {\bf p}) & = & 4 ({\bf p} \times {\bf p}')^2 \\
FA_{4}({\bf p}', {\bf p}) & = & 4 ({\bf p}' + {\bf p})^2 \\
FA_{5}({\bf p}', {\bf p}) & = & 4 ({\bf p}' - {\bf p})^2 \\
FA_{6}({\bf p}', {\bf p}) & = & 4 (p'^2 - p^2)^2 
\end{eqnarray}
\noindent The non-zero coefficients $A_{ij}({\bf p}', {\bf p})$  are:
\begin{eqnarray}
A_{11}({\bf p}', {\bf p}) & = & 4 \\
A_{22}({\bf p}', {\bf p}) & = & 12 \\
A_{24}({\bf p}', {\bf p}) & = & A_{42}({\bf p}', {\bf p}) = FA_{3}({\bf p}', {\bf p}) \\
A_{25}({\bf p}', {\bf p}) & = & A_{52}({\bf p}', {\bf p}) = FA_{4}({\bf p}', {\bf p}) \\
A_{26}({\bf p}', {\bf p}) & = & A_{62}({\bf p}', {\bf p}) = FA_{5}({\bf p}', {\bf p}) \\
A_{33}({\bf p}', {\bf p}) & = & - 2 FA_{3}({\bf p}', {\bf p}) \\ 
A_{44}({\bf p}', {\bf p}) & = & \frac{1}{4} A_{24}^2({\bf p}', {\bf p}) \\
A_{55}({\bf p}', {\bf p}) & = & \frac{1}{4} A_{25}^2({\bf p}', {\bf p}) \\
A_{56}({\bf p}', {\bf p}) & = & A_{65}({\bf p}', {\bf p}) = FA_{6}({\bf p}', {\bf p})
\\
A_{66}({\bf p}', {\bf p}) & = & \frac{1}{4} A_{26}^2({\bf p}', {\bf p}) 
\end{eqnarray}
All other $A_{ij}({\bf p}', {\bf p}) = 0$.

The non-vanishing coefficients $B_{ikj}({\bf p}', {\bf p}'', {\bf p})$ 
can be expressed by means of the following 25 functions $FB$:
\begin{eqnarray}
FB_{3a} ({\bf p}', {\bf p}'', {\bf p}) & = & 4 ({\bf p} \times {\bf p}'')^2 \\
FB_{3b} ({\bf p}', {\bf p}'', {\bf p}) & = & 4 ({\bf p}'' \times {\bf p}')^2 \\
FB_{3c} ({\bf p}', {\bf p}'', {\bf p}) & = & 4 ({\bf p} \times {\bf p}')^2 
\end{eqnarray}
\begin{eqnarray}
FB_{4a} ({\bf p}', {\bf p}'', {\bf p}) & = & 4 ({\bf p}'' + {\bf p})^2 \\
FB_{4b} ({\bf p}', {\bf p}'', {\bf p}) & = & 4 ({\bf p}' + {\bf p}'')^2 \\
FB_{4c} ({\bf p}', {\bf p}'', {\bf p}) & = & 4 ({\bf p}' + {\bf p})^2 
\end{eqnarray}
\begin{eqnarray}
FB_{5a} ({\bf p}', {\bf p}'', {\bf p}) & = & 4 ({\bf p}'' - {\bf p})^2 \\
FB_{5b} ({\bf p}', {\bf p}'', {\bf p}) & = & 4 ({\bf p}' - {\bf p}'')^2 \\
FB_{5c} ({\bf p}', {\bf p}'', {\bf p}) & = & 4 ({\bf p}' - {\bf p})^2 
\end{eqnarray}
\begin{eqnarray}
FB_{6a} ({\bf p}', {\bf p}'', {\bf p}) & = & - 8 ({\bf p}'' \times {\bf p}') \cdot ({\bf p} \times {\bf p}'') \\
FB_{6b} ({\bf p}', {\bf p}'', {\bf p}) & = & - 8 ({\bf p} \times {\bf p}') \cdot ({\bf p}'' \times {\bf p}') \\
FB_{6c} ({\bf p}', {\bf p}'', {\bf p}) & = & - 8 ({\bf p} \times {\bf p}') \cdot ({\bf p} \times {\bf p}'') 
\end{eqnarray}
\begin{eqnarray}
FB_{7}  ({\bf p}', {\bf p}'', {\bf p}) & = & 4 \{({\bf p} \times {\bf p}') \cdot {\bf p}''\}^2 
\end{eqnarray}
\begin{eqnarray}
FB_{8a} ({\bf p}', {\bf p}'', {\bf p}) & = & 2 \{({\bf p}' + {\bf p}'') \cdot ({\bf p}'' + {\bf p})\} \\
FB_{8b} ({\bf p}', {\bf p}'', {\bf p}) & = & 2 \{({\bf p}' + {\bf p}) \cdot ({\bf p}' + {\bf p}'')\} \\
FB_{8c} ({\bf p}', {\bf p}'', {\bf p}) & = & 2 \{({\bf p}' + {\bf p}) \cdot ({\bf p}'' + {\bf p})\} 
\end{eqnarray}
\begin{eqnarray}
FB_{9a} ({\bf p}', {\bf p}'', {\bf p}) & = & 2 \{({\bf p}' - {\bf p}'') \cdot ({\bf p}'' - {\bf p})\} \\
FB_{9b} ({\bf p}', {\bf p}'', {\bf p}) & = & 2 \{({\bf p}' - {\bf p}) \cdot ({\bf p}' - {\bf p}'')\} \\
FB_{9c} ({\bf p}', {\bf p}'', {\bf p}) & = & 2 \{({\bf p}' - {\bf p}) \cdot ({\bf p}'' - {\bf p})\} 
\end{eqnarray}
\begin{eqnarray}
FB_{10a}({\bf p}', {\bf p}'', {\bf p}) & = & 2 \{({\bf p}' + {\bf p}'') \cdot ({\bf p}'' - {\bf p})\} \\
FB_{10b}({\bf p}', {\bf p}'', {\bf p}) & = & 2 \{({\bf p}' + {\bf p}) \cdot ({\bf p}' - {\bf p}'')\} \\
FB_{10c}({\bf p}', {\bf p}'', {\bf p}) & = & 2 \{({\bf p}' - {\bf p}) \cdot ({\bf p}'' + {\bf p})\} 
\end{eqnarray}
\begin{eqnarray}
FB_{11a}({\bf p}', {\bf p}'', {\bf p}) & = & 2 \{({\bf p}' - {\bf p}'') \cdot ({\bf p}'' + {\bf p})\} \\
FB_{11b}({\bf p}', {\bf p}'', {\bf p}) & = & 2 \{({\bf p}' - {\bf p}) \cdot ({\bf p}' + {\bf p}'')\} \\
FB_{11c}({\bf p}', {\bf p}'', {\bf p}) & = & 2 \{({\bf p}' + {\bf p}) \cdot ({\bf p}'' - {\bf p})\} 
\end{eqnarray}

\noindent The non-zero $B_{ikj}({\bf p}', {\bf p}'', {\bf p})$: 
\begin{eqnarray}
B_{122}({\bf p}', {\bf p}'', {\bf p}) & = & B_{212}({\bf p}', {\bf p}'', {\bf p}) = B_{221}({\bf p}', {\bf p}'', {\bf p}) = 12 \\
B_{124}({\bf p}', {\bf p}'', {\bf p}) & = & B_{214}({\bf p}', {\bf p}'', {\bf p}) = FB_{3a}({\bf p}', {\bf p}'', {\bf p}) \\
B_{125}({\bf p}', {\bf p}'', {\bf p}) & = & B_{215}({\bf p}', {\bf p}'', {\bf p}) = FB_{4a}({\bf p}', {\bf p}'', {\bf p}) \\
B_{126}({\bf p}', {\bf p}'', {\bf p}) & = & B_{216}({\bf p}', {\bf p}'', {\bf p}) = FB_{5a}({\bf p}', {\bf p}'', {\bf p}) \\
B_{133}({\bf p}', {\bf p}'', {\bf p}) & = & B_{233}({\bf p}', {\bf p}'', {\bf p}) = FB_{6a}({\bf p}', {\bf p}'', {\bf p}) \\
B_{142}({\bf p}', {\bf p}'', {\bf p}) & = & B_{241}({\bf p}', {\bf p}'', {\bf p}) = FB_{3b}({\bf p}', {\bf p}'', {\bf p}) \\
B_{144}({\bf p}', {\bf p}'', {\bf p}) & = & \frac{1}{16} FB_{6a}({\bf p}', {\bf p}'', {\bf p})^2 
\end{eqnarray}
\begin{eqnarray}
B_{145}({\bf p}', {\bf p}'', {\bf p}) & = & B_{146}({\bf p}', {\bf p}'', {\bf p}) 
                                        =   B_{154}({\bf p}', {\bf p}'', {\bf p}) 
                                        =   B_{164}({\bf p}', {\bf p}'', {\bf p}) \nonumber\\
                                      & = & B_{415}({\bf p}', {\bf p}'', {\bf p}) 
                                        =   B_{416}({\bf p}', {\bf p}'', {\bf p}) 
                                        =   B_{514}({\bf p}', {\bf p}'', {\bf p}) \nonumber\\
                                      & = & B_{614}({\bf p}', {\bf p}'', {\bf p}) 
                                        =   B_{451}({\bf p}', {\bf p}'', {\bf p}) 
                                        =   B_{461}({\bf p}', {\bf p}'', {\bf p}) \nonumber\\
                                      & = & B_{541}({\bf p}', {\bf p}'', {\bf p}) 
                                        =   B_{641}({\bf p}', {\bf p}'', {\bf p}) 
                                        =   FB_{7}({\bf p}', {\bf p}'', {\bf p})
\end{eqnarray}
\begin{eqnarray}
B_{152}({\bf p}', {\bf p}'', {\bf p}) & = & B_{251}({\bf p}', {\bf p}'', {\bf p}) = FB_{4b}({\bf p}', {\bf p}'', {\bf p}) \\
B_{155}({\bf p}', {\bf p}'', {\bf p}) & = & FB_{8a} ({\bf p}', {\bf p}'', {\bf p})^2 \\
B_{156}({\bf p}', {\bf p}'', {\bf p}) & = & FB_{10a}({\bf p}', {\bf p}'', {\bf p})^2 \\
B_{162}({\bf p}', {\bf p}'', {\bf p}) & = & B_{261}({\bf p}', {\bf p}'', {\bf p}) = FB_{5b}({\bf p}', {\bf p}'', {\bf p}) \\
B_{165}({\bf p}', {\bf p}'', {\bf p}) & = & FB_{11a}({\bf p}', {\bf p}'', {\bf p})^2 \\
B_{166}({\bf p}', {\bf p}'', {\bf p}) & = & FB_{9a} ({\bf p}', {\bf p}'', {\bf p})^2 \\
B_{313}({\bf p}', {\bf p}'', {\bf p}) & = & B_{323}({\bf p}', {\bf p}'', {\bf p}) = FB_{6c} ({\bf p}', {\bf p}'', {\bf p}) 
\end{eqnarray}
\begin{eqnarray}
B_{412}({\bf p}', {\bf p}'', {\bf p}) & = & B_{421}({\bf p}', {\bf p}'', {\bf p}) = FB_{3c} ({\bf p}', {\bf p}'', {\bf p}) \\
B_{414}({\bf p}', {\bf p}'', {\bf p}) & = & \frac{1}{16} FB_{6c}({\bf p}', {\bf p}'', {\bf p})^2 \\
B_{512}({\bf p}', {\bf p}'', {\bf p}) & = & B_{521}({\bf p}', {\bf p}'', {\bf p}) = FB_{4c} ({\bf p}', {\bf p}'', {\bf p}) \\
B_{515}({\bf p}', {\bf p}'', {\bf p}) & = & FB_{8c} ({\bf p}', {\bf p}'', {\bf p})^2 \\
B_{516}({\bf p}', {\bf p}'', {\bf p}) & = & FB_{11c}({\bf p}', {\bf p}'', {\bf p})^2 \\
B_{612}({\bf p}', {\bf p}'', {\bf p}) & = & B_{621}({\bf p}', {\bf p}'', {\bf p}) = FB_{5c} ({\bf p}', {\bf p}'', {\bf p}) 
\end{eqnarray}
\begin{eqnarray}
B_{615}({\bf p}', {\bf p}'', {\bf p}) & = & FB_{10c}({\bf p}', {\bf p}'', {\bf p})^2 \\
B_{616}({\bf p}', {\bf p}'', {\bf p}) & = & FB_{9c} ({\bf p}', {\bf p}'', {\bf p})^2 \\
B_{331}({\bf p}', {\bf p}'', {\bf p}) & = & B_{332}({\bf p}', {\bf p}'', {\bf p}) = FB_{6b} ({\bf p}', {\bf p}'', {\bf p}) \\
B_{441}({\bf p}', {\bf p}'', {\bf p}) & = & \frac{1}{16} FB_{6b}({\bf p}', {\bf p}'', {\bf p})^2 \\
B_{551}({\bf p}', {\bf p}'', {\bf p}) & = & FB_{8b} ({\bf p}', {\bf p}'', {\bf p})^2 \\
B_{561}({\bf p}', {\bf p}'', {\bf p}) & = & FB_{10b}({\bf p}', {\bf p}'', {\bf p})^2 \\
B_{651}({\bf p}', {\bf p}'', {\bf p}) & = & FB_{11b}({\bf p}', {\bf p}'', {\bf p})^2 \\
B_{661}({\bf p}', {\bf p}'', {\bf p}) & = & FB_{9b} ({\bf p}', {\bf p}'', {\bf p})^2 \\
B_{111}({\bf p}', {\bf p}'', {\bf p}) & = & 4 
\end{eqnarray}
\begin{eqnarray}
B_{244}({\bf p}', {\bf p}'', {\bf p}) & = & - p''^2 FB_{7}({\bf p}', {\bf p}'', {\bf p}) \\
B_{245}({\bf p}', {\bf p}'', {\bf p}) & = & - \frac{1}{4} FB_{3b}({\bf p}', {\bf p}'', {\bf p}) FB_{4a}({\bf p}', {\bf p}'', {\bf p}) \nonumber\\
&&                                          + FB_{7}({\bf p}', {\bf p}'', {\bf p}) \\
B_{246}({\bf p}', {\bf p}'', {\bf p}) & = & - \frac{1}{4} FB_{3b}({\bf p}', {\bf p}'', {\bf p}) FB_{5a}({\bf p}', {\bf p}'', {\bf p}) \nonumber\\
&&                                          + FB_{7}({\bf p}', {\bf p}'', {\bf p}) \\
B_{254}({\bf p}', {\bf p}'', {\bf p}) & = & - \frac{1}{4} FB_{3a}({\bf p}', {\bf p}'', {\bf p}) FB_{4b}({\bf p}', {\bf p}'', {\bf p}) \nonumber\\
&&                                          + FB_{7}({\bf p}', {\bf p}'', {\bf p}) \\
B_{255}({\bf p}', {\bf p}'', {\bf p}) & = & - \frac{1}{4} FB_{4b}({\bf p}', {\bf p}'', {\bf p}) FB_{4a}({\bf p}', {\bf p}'', {\bf p}) \nonumber\\
&&                                          + FB_{8a}({\bf p}', {\bf p}'', {\bf p})^2
\end{eqnarray}
\begin{eqnarray}
B_{256}({\bf p}', {\bf p}'', {\bf p}) & = & - \frac{1}{4} FB_{4b}({\bf p}', {\bf p}'', {\bf p}) FB_{5a}({\bf p}', {\bf p}'', {\bf p}) \nonumber\\
&&                                          + FB_{10a}({\bf p}', {\bf p}'', {\bf p})^2 \\
B_{264}({\bf p}', {\bf p}'', {\bf p}) & = & - \frac{1}{4} FB_{3a}({\bf p}', {\bf p}'', {\bf p}) FB_{5b}({\bf p}', {\bf p}'', {\bf p}) \nonumber\\
&&                                          + FB_{7}({\bf p}', {\bf p}'', {\bf p})
\end{eqnarray}
\begin{eqnarray}
B_{265}({\bf p}', {\bf p}'', {\bf p}) & = & - \frac{1}{4} FB_{5b}({\bf p}', {\bf p}'', {\bf p}) FB_{4a}({\bf p}', {\bf p}'', {\bf p}) \nonumber\\
&&                                          + FB_{11a}({\bf p}', {\bf p}'', {\bf p})^2 \\
B_{266}({\bf p}', {\bf p}'', {\bf p}) & = & - \frac{1}{4} FB_{5b}({\bf p}', {\bf p}'', {\bf p}) FB_{5a}({\bf p}', {\bf p}'', {\bf p}) \nonumber\\
&&                                          + FB_{9a}({\bf p}', {\bf p}'', {\bf p})^2 
\end{eqnarray}
\begin{eqnarray}
B_{424}({\bf p}', {\bf p}'', {\bf p}) & = & - p^2 FB_{7}({\bf p}', {\bf p}'', {\bf p}) \\
B_{425}({\bf p}', {\bf p}'', {\bf p}) & = & - \frac{1}{4} FB_{3c}({\bf p}', {\bf p}'', {\bf p}) FB_{4a}({\bf p}', {\bf p}'', {\bf p}) \nonumber\\
&&                                          + FB_{7}({\bf p}', {\bf p}'', {\bf p}) \\
B_{426}({\bf p}', {\bf p}'', {\bf p}) & = & - \frac{1}{4} FB_{3c}({\bf p}', {\bf p}'', {\bf p}) FB_{5a}({\bf p}', {\bf p}'', {\bf p}) \nonumber\\
&&                                          + FB_{7}({\bf p}', {\bf p}'', {\bf p}) \\
B_{524}({\bf p}', {\bf p}'', {\bf p}) & = & - \frac{1}{4} FB_{3a}({\bf p}', {\bf p}'', {\bf p}) FB_{4c}({\bf p}', {\bf p}'', {\bf p}) \nonumber\\
&&                                          + FB_{7}({\bf p}', {\bf p}'', {\bf p}) \\
B_{525}({\bf p}', {\bf p}'', {\bf p}) & = & - \frac{1}{4} FB_{4a}({\bf p}', {\bf p}'', {\bf p}) FB_{4c}({\bf p}', {\bf p}'', {\bf p}) \nonumber\\
&&                                          + FB_{8c}({\bf p}', {\bf p}'', {\bf p})^2
\end{eqnarray}
\begin{eqnarray}
B_{526}({\bf p}', {\bf p}'', {\bf p}) & = & - \frac{1}{4} FB_{5a}({\bf p}', {\bf p}'', {\bf p}) FB_{4c}({\bf p}', {\bf p}'', {\bf p}) \nonumber\\
&&                                          + FB_{11c}({\bf p}', {\bf p}'', {\bf p})^2 \\
B_{624}({\bf p}', {\bf p}'', {\bf p}) & = & - \frac{1}{4} FB_{3a}({\bf p}', {\bf p}'', {\bf p}) FB_{5c}({\bf p}', {\bf p}'', {\bf p}) \nonumber\\
&&                                          + FB_{7}({\bf p}', {\bf p}'', {\bf p}) \\
B_{625}({\bf p}', {\bf p}'', {\bf p}) & = & - \frac{1}{4} FB_{4a}({\bf p}', {\bf p}'', {\bf p}) FB_{5c}({\bf p}', {\bf p}'', {\bf p}) \nonumber\\
&&                                          + FB_{10c}({\bf p}', {\bf p}'', {\bf p})^2 \\
B_{626}({\bf p}', {\bf p}'', {\bf p}) & = & - \frac{1}{4} FB_{5a}({\bf p}', {\bf p}'', {\bf p}) FB_{5c}({\bf p}', {\bf p}'', {\bf p}) \nonumber\\
&&                                          + FB_{9c}({\bf p}', {\bf p}'', {\bf p})^2 \\
B_{442}({\bf p}', {\bf p}'', {\bf p}) & = & - p'^2 FB_{7}({\bf p}', {\bf p}'', {\bf p}) \\
B_{452}({\bf p}', {\bf p}'', {\bf p}) & = & - \frac{1}{4} FB_{3c}({\bf p}', {\bf p}'', {\bf p}) FB_{4b}({\bf p}', {\bf p}'', {\bf p}) \nonumber\\
&&                                          + FB_{7}({\bf p}', {\bf p}'', {\bf p})
\end{eqnarray}
\begin{eqnarray}
B_{462}({\bf p}', {\bf p}'', {\bf p}) & = & - \frac{1}{4} FB_{3c}({\bf p}', {\bf p}'', {\bf p}) FB_{5b}({\bf p}', {\bf p}'', {\bf p}) \nonumber\\
&&                                          + FB_{7}({\bf p}', {\bf p}'', {\bf p}) \\
B_{542}({\bf p}', {\bf p}'', {\bf p}) & = & - \frac{1}{4} FB_{3b}({\bf p}', {\bf p}'', {\bf p}) FB_{4c}({\bf p}', {\bf p}'', {\bf p}) \nonumber\\
&&                                          + FB_{7}({\bf p}', {\bf p}'', {\bf p}) \\
B_{552}({\bf p}', {\bf p}'', {\bf p}) & = & - \frac{1}{4} FB_{4c}({\bf p}', {\bf p}'', {\bf p}) FB_{4b}({\bf p}', {\bf p}'', {\bf p}) \nonumber\\
&&                                          + FB_{8b}({\bf p}', {\bf p}'', {\bf p})^2 \\
B_{562}({\bf p}', {\bf p}'', {\bf p}) & = & - \frac{1}{4} FB_{4c}({\bf p}', {\bf p}'', {\bf p}) FB_{5b}({\bf p}', {\bf p}'', {\bf p}) \nonumber\\
&&                                          + FB_{10b}({\bf p}', {\bf p}'', {\bf p})^2 
\end{eqnarray}
\begin{eqnarray}
B_{642}({\bf p}', {\bf p}'', {\bf p}) & = & - \frac{1}{4} FB_{3b}({\bf p}', {\bf p}'', {\bf p}) FB_{5c}({\bf p}', {\bf p}'', {\bf p}) \nonumber\\
&&                                          + FB_{7}({\bf p}', {\bf p}'', {\bf p}) \\
B_{652}({\bf p}', {\bf p}'', {\bf p}) & = & - \frac{1}{4} FB_{5c}({\bf p}', {\bf p}'', {\bf p}) FB_{4b}({\bf p}', {\bf p}'', {\bf p}) \nonumber\\
&&                                          + FB_{11b}({\bf p}', {\bf p}'', {\bf p})^2 \\
B_{662}({\bf p}', {\bf p}'', {\bf p}) & = & - \frac{1}{4} FB_{5c}({\bf p}', {\bf p}'', {\bf p}) FB_{5b}({\bf p}', {\bf p}'', {\bf p}) \nonumber\\
&&                                          + FB_{9b}({\bf p}', {\bf p}'', {\bf p})^2
\end{eqnarray}
\begin{eqnarray}
B_{224}({\bf p}', {\bf p}'', {\bf p}) & = & -2 FB_{3a}({\bf p}', {\bf p}'', {\bf p}) \\
B_{225}({\bf p}', {\bf p}'', {\bf p}) & = & -2 FB_{4a}({\bf p}', {\bf p}'', {\bf p}) \\
B_{226}({\bf p}', {\bf p}'', {\bf p}) & = & -2 FB_{5a}({\bf p}', {\bf p}'', {\bf p}) \\
B_{242}({\bf p}', {\bf p}'', {\bf p}) & = & -2 FB_{3b}({\bf p}', {\bf p}'', {\bf p}) \\
B_{252}({\bf p}', {\bf p}'', {\bf p}) & = & -2 FB_{4b}({\bf p}', {\bf p}'', {\bf p})
\end{eqnarray}
\begin{eqnarray}
B_{262}({\bf p}', {\bf p}'', {\bf p}) & = & -2 FB_{5b}({\bf p}', {\bf p}'', {\bf p}) \\
B_{422}({\bf p}', {\bf p}'', {\bf p}) & = & -2 FB_{3c}({\bf p}', {\bf p}'', {\bf p}) \\
B_{522}({\bf p}', {\bf p}'', {\bf p}) & = & -2 FB_{4c}({\bf p}', {\bf p}'', {\bf p}) \\
B_{622}({\bf p}', {\bf p}'', {\bf p}) & = & -2 FB_{5c}({\bf p}', {\bf p}'', {\bf p}) \\
B_{222}({\bf p}', {\bf p}'', {\bf p}) & = & - 24
\end{eqnarray}
\begin{eqnarray}
B_{344}({\bf p}', {\bf p}'', {\bf p}) & = & \frac{1}{4} FB_{6a}({\bf p}', {\bf p}'', {\bf p}) 
                                                        FB_{7}({\bf p}', {\bf p}'', {\bf p}) \\
B_{345}({\bf p}', {\bf p}'', {\bf p}) & = & B_{435}({\bf p}', {\bf p}'', {\bf p}) 
                                        =   - 2 \{({\bf p}'' + {\bf p}) \cdot {\bf p}'\} FB_{7}({\bf p}', {\bf p}'', {\bf p}) \\
B_{346}({\bf p}', {\bf p}'', {\bf p}) & = & B_{436}({\bf p}', {\bf p}'', {\bf p}) 
                                        =   - 2 \{({\bf p}'' - {\bf p}) \cdot {\bf p}'\} FB_{7}({\bf p}', {\bf p}'', {\bf p})
\end{eqnarray}
\begin{eqnarray}
B_{355}({\bf p}', {\bf p}'', {\bf p}) & = & \frac{1}{2} \left\lbrace  2 FB_{3c}({\bf p}', {\bf p}'', {\bf p})
                                                                     - FB_{6b}({\bf p}', {\bf p}'', {\bf p}) \right.\nonumber\\
&&                                                       \left.      - FB_{6c}({\bf p}', {\bf p}'', {\bf p})\right\rbrace 
                                                                       FB_{8a}({\bf p}', {\bf p}'', {\bf p}) \quad 
\end{eqnarray}
\begin{eqnarray}
B_{356}({\bf p}', {\bf p}'', {\bf p}) & = & -\frac{1}{2} \left\lbrace  2 FB_{3c}({\bf p}', {\bf p}'', {\bf p})
                                                                     + FB_{6b}({\bf p}', {\bf p}'', {\bf p}) \right.\nonumber\\
&&                                                       \left.      - FB_{6c}({\bf p}', {\bf p}'', {\bf p})\right\rbrace 
                                                                       FB_{10a}({\bf p}', {\bf p}'', {\bf p}) \quad \\
B_{365}({\bf p}', {\bf p}'', {\bf p}) & = & \frac{1}{2}\left\lbrace  2 FB_{3c}({\bf p}', {\bf p}'', {\bf p})
                                                                    - FB_{6b}({\bf p}', {\bf p}'', {\bf p}) \right.\nonumber\\
&&                                                      \left.      + FB_{6c}({\bf p}', {\bf p}'', {\bf p})\right\rbrace 
                                                                      FB_{11a}({\bf p}', {\bf p}'', {\bf p}) \quad \\
B_{366}({\bf p}', {\bf p}'', {\bf p}) & = & -\frac{1}{2} \left\lbrace  2 FB_{3c}({\bf p}', {\bf p}'', {\bf p})
                                                                     + FB_{6b}({\bf p}', {\bf p}'', {\bf p}) \right.\nonumber\\
&&                                                       \left.      + FB_{6c}({\bf p}', {\bf p}'', {\bf p})\right\rbrace 
                                                                       FB_{9a}({\bf p}', {\bf p}'', {\bf p}) \qquad \\
B_{434}({\bf p}', {\bf p}'', {\bf p}) & = & \frac{1}{4} FB_{6c}({\bf p}', {\bf p}'', {\bf p}) 
                                                        FB_{7}({\bf p}', {\bf p}'', {\bf p}) 
\end{eqnarray}
\begin{eqnarray}
B_{534}({\bf p}', {\bf p}'', {\bf p}) & = & B_{543}({\bf p}', {\bf p}'', {\bf p}) 
                                        =   2 \{({\bf p}' + {\bf p}) \cdot {\bf p}''\} FB_{7}({\bf p}', {\bf p}'', {\bf p}) \\
B_{634}({\bf p}', {\bf p}'', {\bf p}) & = & - B_{643}({\bf p}', {\bf p}'', {\bf p}) 
                                        =   2 \{({\bf p}' - {\bf p}) \cdot {\bf p}''\} FB_{7}({\bf p}', {\bf p}'', {\bf p}) \\
B_{535}({\bf p}', {\bf p}'', {\bf p}) & = & -\frac{1}{2} \left\lbrace  2 FB_{3b}({\bf p}', {\bf p}'', {\bf p}) 
                                                                     + FB_{6a}({\bf p}', {\bf p}'', {\bf p}) \right.\nonumber\\
&&                                                       \left.      - FB_{6b}({\bf p}', {\bf p}'', {\bf p})\right\rbrace 
                                                                       FB_{8c}({\bf p}', {\bf p}'', {\bf p}) \quad \\
B_{536}({\bf p}', {\bf p}'', {\bf p}) & = & -\frac{1}{2} \left\lbrace  2 FB_{3b}({\bf p}', {\bf p}'', {\bf p})
                                                                     + FB_{6a}({\bf p}', {\bf p}'', {\bf p}) \right.\nonumber\\
&&                                                       \left.      + FB_{6b}({\bf p}', {\bf p}'', {\bf p})\right\rbrace 
                                                                       FB_{11c}({\bf p}', {\bf p}'', {\bf p}) \quad 
\end{eqnarray}
\begin{eqnarray}
B_{635}({\bf p}', {\bf p}'', {\bf p}) & = & -\frac{1}{2} \left\lbrace  2 FB_{3b}({\bf p}', {\bf p}'', {\bf p})
                                                                     - FB_{6a}({\bf p}', {\bf p}'', {\bf p}) \right.\nonumber\\
&&                                                       \left.      - FB_{6b}({\bf p}', {\bf p}'', {\bf p})\right\rbrace 
                                                                       FB_{10c}({\bf p}', {\bf p}'', {\bf p}) \qquad \\
B_{636}({\bf p}', {\bf p}'', {\bf p}) & = & -\frac{1}{2} \left\lbrace  2 FB_{3b}({\bf p}', {\bf p}'', {\bf p})
                                                                     - FB_{6a}({\bf p}', {\bf p}'', {\bf p}) \right.\nonumber\\
&&                                                       \left.      + FB_{6b}({\bf p}', {\bf p}'', {\bf p})\right\rbrace 
                                                                       FB_{9c}({\bf p}', {\bf p}'', {\bf p}) \quad 
\end{eqnarray}
\begin{eqnarray}
B_{443}({\bf p}', {\bf p}'', {\bf p}) & = & \frac{1}{4} FB_{6b}({\bf p}', {\bf p}'', {\bf p}) 
                                                        FB_{7}({\bf p}', {\bf p}'', {\bf p}) \\
B_{453}({\bf p}', {\bf p}'', {\bf p}) & = & - B_{354}({\bf p}', {\bf p}'', {\bf p}) 
                                        =   2 \{({\bf p}' + {\bf p}'') \cdot {\bf p}\} FB_{7}({\bf p}', {\bf p}'', {\bf p}) \\
B_{463}({\bf p}', {\bf p}'', {\bf p}) & = & B_{364}({\bf p}', {\bf p}'', {\bf p}) 
                                        =   - 2 \{({\bf p}' - {\bf p}'') \cdot {\bf p}\} FB_{7}({\bf p}', {\bf p}'', {\bf p}) \qquad \\
B_{553}({\bf p}', {\bf p}'', {\bf p}) & = & -\frac{1}{2} \left\lbrace  2 FB_{3a}({\bf p}', {\bf p}'', {\bf p})
                                                                     - FB_{6c}({\bf p}', {\bf p}'', {\bf p}) \right.\nonumber\\
&&                                                       \left.      + FB_{6a}({\bf p}', {\bf p}'', {\bf p})\right\rbrace 
                                                                       FB_{8b}({\bf p}', {\bf p}'', {\bf p})
\end{eqnarray}
\begin{eqnarray}
B_{563}({\bf p}', {\bf p}'', {\bf p}) & = & \frac{1}{2} \left\lbrace  2 FB_{3a}({\bf p}', {\bf p}'', {\bf p})
                                                                     + FB_{6c}({\bf p}', {\bf p}'', {\bf p}) \right.\nonumber\\
&&                                                       \left.      + FB_{6a}({\bf p}', {\bf p}'', {\bf p})\right\rbrace 
                                                                       FB_{10b}({\bf p}', {\bf p}'', {\bf p}) \\
B_{653}({\bf p}', {\bf p}'', {\bf p}) & = & \frac{1}{2} \left\lbrace  2 FB_{3a}({\bf p}', {\bf p}'', {\bf p})
                                                                     - FB_{6c}({\bf p}', {\bf p}'', {\bf p}) \right.\nonumber\\
&&                                                       \left.      - FB_{6a}({\bf p}', {\bf p}'', {\bf p})\right\rbrace 
                                                                       FB_{11b}({\bf p}', {\bf p}'', {\bf p}) \\
B_{663}({\bf p}', {\bf p}'', {\bf p}) & = & -\frac{1}{2} \left\lbrace  2 FB_{3a}({\bf p}', {\bf p}'', {\bf p})
                                                                     + FB_{6c}({\bf p}', {\bf p}'', {\bf p}) \right.\nonumber\\
&&                                                       \left.      - FB_{6a}({\bf p}', {\bf p}'', {\bf p})\right\rbrace 
                                                                       FB_{9b}({\bf p}', {\bf p}'', {\bf p})
\end{eqnarray}
\begin{eqnarray}
B_{334}({\bf p}', {\bf p}'', {\bf p}) & = & - \frac{1}{8} FB_{6a}({\bf p}', {\bf p}'', {\bf p}) 
                                                          FB_{6c}({\bf p}', {\bf p}'', {\bf p}) \\
B_{335}({\bf p}', {\bf p}'', {\bf p}) & = & B_{353}({\bf p}', {\bf p}'', {\bf p}) = B_{633}({\bf p}', {\bf p}'', {\bf p}) \nonumber\\
                                      & = & B_{333}({\bf p}', {\bf p}'', {\bf p}) = 2 FB_{7}({\bf p}', {\bf p}'', {\bf p}) \\
B_{336}({\bf p}', {\bf p}'', {\bf p}) & = & B_{363}({\bf p}', {\bf p}'', {\bf p}) = B_{533}({\bf p}', {\bf p}'', {\bf p}) \nonumber\\
                                      & = & - 2 FB_{7}({\bf p}', {\bf p}'', {\bf p}) \\
B_{343}({\bf p}', {\bf p}'', {\bf p}) & = & - \frac{1}{8} FB_{6b}({\bf p}', {\bf p}'', {\bf p}) 
                                                          FB_{6a}({\bf p}', {\bf p}'', {\bf p}) \\
B_{433}({\bf p}', {\bf p}'', {\bf p}) & = & - \frac{1}{8} FB_{6c}({\bf p}', {\bf p}'', {\bf p}) 
                                                          FB_{6b}({\bf p}', {\bf p}'', {\bf p}) 
\end{eqnarray}
\begin{eqnarray}
B_{455}({\bf p}', {\bf p}'', {\bf p}) & = & -\frac{1}{16} \left\lbrace  2 FB_{3c}({\bf p}', {\bf p}'', {\bf p})
                                                                      - FB_{6b}({\bf p}', {\bf p}'', {\bf p}) \right.\nonumber\\
&&                                                       \left.      - FB_{6c}({\bf p}', {\bf p}'', {\bf p})\right\rbrace^2 \\ 
B_{456}({\bf p}', {\bf p}'', {\bf p}) & = & -\frac{1}{16} \left\lbrace  2 FB_{3c}({\bf p}', {\bf p}'', {\bf p})
                                                                      + FB_{6b}({\bf p}', {\bf p}'', {\bf p}) \right.\nonumber\\
&&                                                       \left.      - FB_{6c}({\bf p}', {\bf p}'', {\bf p})\right\rbrace^2 \\ 
B_{465}({\bf p}', {\bf p}'', {\bf p}) & = & -\frac{1}{16} \left\lbrace  2 FB_{3c}({\bf p}', {\bf p}'', {\bf p})
                                                                      - FB_{6b}({\bf p}', {\bf p}'', {\bf p}) \right.\nonumber\\
&&                                                       \left.      + FB_{6c}({\bf p}', {\bf p}'', {\bf p})\right\rbrace^2 \\ 
B_{466}({\bf p}', {\bf p}'', {\bf p}) & = & -\frac{1}{16} \left\lbrace  2 FB_{3c}({\bf p}', {\bf p}'', {\bf p})
                                                                      + FB_{6b}({\bf p}', {\bf p}'', {\bf p}) \right.\nonumber\\
&&                                                       \left.      + FB_{6c}({\bf p}', {\bf p}'', {\bf p})\right\rbrace^2 
\end{eqnarray}
\begin{eqnarray}
B_{545}({\bf p}', {\bf p}'', {\bf p}) & = & -\frac{1}{16} \left\lbrace  2 FB_{3b}({\bf p}', {\bf p}'', {\bf p}) 
                                                                      + FB_{6a}({\bf p}', {\bf p}'', {\bf p}) \right.\nonumber\\
&&                                                       \left.      - FB_{6b}({\bf p}', {\bf p}'', {\bf p})\right\rbrace^2 \\ 
B_{546}({\bf p}', {\bf p}'', {\bf p}) & = & -\frac{1}{16} \left\lbrace  2 FB_{3b}({\bf p}', {\bf p}'', {\bf p})
                                                                      + FB_{6a}({\bf p}', {\bf p}'', {\bf p}) \right.\nonumber\\
&&                                                       \left.      + FB_{6b}({\bf p}', {\bf p}'', {\bf p})\right\rbrace^2 \\
B_{645}({\bf p}', {\bf p}'', {\bf p}) & = & -\frac{1}{16} \left\lbrace  2 FB_{3b}({\bf p}', {\bf p}'', {\bf p})
                                                                      - FB_{6a}({\bf p}', {\bf p}'', {\bf p}) \right.\nonumber\\
&&                                                       \left.      - FB_{6b}({\bf p}', {\bf p}'', {\bf p})\right\rbrace^2 \\ 
B_{646}({\bf p}', {\bf p}'', {\bf p}) & = & -\frac{1}{16} \left\lbrace  2 FB_{3b}({\bf p}', {\bf p}'', {\bf p})
                                                                      - FB_{6a}({\bf p}', {\bf p}'', {\bf p}) \right.\nonumber\\
&&                                                       \left.      + FB_{6b}({\bf p}', {\bf p}'', {\bf p})\right\rbrace^2
\end{eqnarray}
\begin{eqnarray}
B_{554}({\bf p}', {\bf p}'', {\bf p}) & = & -\frac{1}{16} \left\lbrace  2 FB_{3a}({\bf p}', {\bf p}'', {\bf p})
                                                                      - FB_{6c}({\bf p}', {\bf p}'', {\bf p}) \right.\nonumber\\
&&                                                       \left.      + FB_{6a}({\bf p}', {\bf p}'', {\bf p})\right\rbrace^2 \\ 
B_{564}({\bf p}', {\bf p}'', {\bf p}) & = & -\frac{1}{16} \left\lbrace  2 FB_{3a}({\bf p}', {\bf p}'', {\bf p})
                                                                      + FB_{6c}({\bf p}', {\bf p}'', {\bf p}) \right.\nonumber\\
&&                                                       \left.      + FB_{6a}({\bf p}', {\bf p}'', {\bf p})\right\rbrace^2 \\ 
B_{654}({\bf p}', {\bf p}'', {\bf p}) & = & -\frac{1}{16} \left\lbrace  2 FB_{3a}({\bf p}', {\bf p}'', {\bf p})
                                                                      - FB_{6c}({\bf p}', {\bf p}'', {\bf p}) \right.\nonumber\\
&&                                                       \left.      - FB_{6a}({\bf p}', {\bf p}'', {\bf p})\right\rbrace^2 \\ 
B_{664}({\bf p}', {\bf p}'', {\bf p}) & = & -\frac{1}{16} \left\lbrace  2 FB_{3a}({\bf p}', {\bf p}'', {\bf p})
                                                                      + FB_{6c}({\bf p}', {\bf p}'', {\bf p}) \right.\nonumber\\
&&                                                       \left.      - FB_{6a}({\bf p}', {\bf p}'', {\bf p})\right\rbrace^2
\end{eqnarray}
\begin{eqnarray}
B_{445}({\bf p}', {\bf p}'', {\bf p}) & = & - \{({\bf p}'' + {\bf p}) \cdot {\bf p}'\}^2 FB_{7}({\bf p}', {\bf p}'', {\bf p}) \\
B_{446}({\bf p}', {\bf p}'', {\bf p}) & = & - \{({\bf p}'' - {\bf p}) \cdot {\bf p}'\}^2 FB_{7}({\bf p}', {\bf p}'', {\bf p}) \\
B_{454}({\bf p}', {\bf p}'', {\bf p}) & = & - \{({\bf p}' + {\bf p}'') \cdot {\bf p}\}^2 FB_{7}({\bf p}', {\bf p}'', {\bf p}) \\
B_{464}({\bf p}', {\bf p}'', {\bf p}) & = & - \{({\bf p}' - {\bf p}'') \cdot {\bf p}\}^2 FB_{7}({\bf p}', {\bf p}'', {\bf p}) 
\end{eqnarray}
\begin{eqnarray}
B_{544}({\bf p}', {\bf p}'', {\bf p}) & = & - \{({\bf p}' + {\bf p}) \cdot {\bf p}''\}^2 FB_{7}({\bf p}', {\bf p}'', {\bf p}) \\
B_{644}({\bf p}', {\bf p}'', {\bf p}) & = & - \{({\bf p}' - {\bf p}) \cdot {\bf p}''\}^2 FB_{7}({\bf p}', {\bf p}'', {\bf p}) \\
B_{444}({\bf p}', {\bf p}'', {\bf p}) & = & - \frac{1}{4} FB_{7}^2({\bf p}', {\bf p}'', {\bf p}) \\
B_{566}({\bf p}', {\bf p}'', {\bf p}) & = & B_{656}({\bf p}', {\bf p}'', {\bf p}) = B_{665}({\bf p}', {\bf p}'', {\bf p}) \nonumber\\
                                      & = & B_{555}({\bf p}', {\bf p}'', {\bf p}) = - 4 FB_{7}({\bf p}', {\bf p}'', {\bf p}) 
\end{eqnarray}

\section{Coefficients for the deuteron}
\label{appendixb}
In this appendix  we present the  expressions $A^ d$  and $B^ d $  given in 
Eqs.~(\ref{eq:2.12}) for the deuteron.
\begin{eqnarray}
A^d_{11} (p) & = &  3\cr  
A^d_{12} (p)&  = &  A^ d_{21} (p) = 0\cr 
A^d_{22} (p)&  = &  \frac{8}{3} p^4
\end{eqnarray}
The coefficients $B^d_{kjj'} ({\bf p},{\bf p}^{\prime})$ are explicitly calculated as
\begin{eqnarray}
B^d_{111} ({\bf p},{\bf p}^{\prime})&=& B^d_{121}({\bf p},{\bf p}^{\prime}) = 3\cr 
B^d_{131} ({\bf p},{\bf p}^{\prime})&=& 0 \cr
B^d_{141} ({\bf p},{\bf p}^{\prime})&=& ({\bf p}\times {\bf p'})^2 \cr
B^d_{151} ({\bf p},{\bf p}^{\prime})&=& ({\bf p'}+ {\bf p})^2 \cr
B^d_{161} ({\bf p},{\bf p}^{\prime})&=& ({\bf p'}- {\bf p})^2 
\end{eqnarray}
\begin{eqnarray}
B^d_{212} ({\bf p},{\bf p}^{\prime}) &=& B^d_{222} ({\bf p},{\bf p}^{\prime})  =
   4({\bf p}\cdot {\bf p'})^2 -\frac{4}{3} p^2 p'^2 \cr
B^d_{232} ({\bf p},{\bf p}^{\prime}) &=& -8 \; {\bf p}\cdot {\bf p'} \; ({\bf p}\times {\bf p'})^2 \cr
B^d_{242} ({\bf p},{\bf p}^{\prime}) &=& -\frac{20}{9} \; p^4 p'^4 -  4({\bf p}\cdot {\bf p'})^4 
+ \frac{56}{9} p^2 p'^2 ({\bf p}\cdot {\bf p'})^2 \cr
B^d_{252} ({\bf p},{\bf p}^{\prime}) &=& \frac{4}{9} \; p^2 p'^2 (p^2 + p'^2) 
- \frac{16}{9} p^2 p'^2 ({\bf p}\cdot {\bf p'})
- \frac{4}{3} (p^2 + p'^2) ({\bf p}\cdot {\bf p'})^2 \cr
B^d_{262} ({\bf p},{\bf p}^{\prime}) &=& \frac{4}{9} \; p^2 p'^2 (p^2 + p'^2) 
+ \frac{16}{9} p^2 p'^2 ({\bf p}\cdot {\bf p'})
- \frac{4}{3} (p^2 + p'^2) ({\bf p}\cdot {\bf p'})^2 
\end{eqnarray}
\begin{eqnarray}
B^d_{211} ({\bf p},{\bf p}^{\prime})&=& B^d_{221}({\bf p},{\bf p}^{\prime})
  = B^ d_{231}({\bf p},{\bf p}^{\prime}) =  0\cr
B^d_{241}({\bf p},{\bf p}^{\prime}) &=&  
    -\frac{4}{3} \; p^2 \; ({\bf p'}\times {\bf p})^2 \cr
B^d_{251}({\bf p},{\bf p}^{\prime})&=&  
  \frac{8}{3} p^4  
+ \frac{16}{3} p^2 \;  ({\bf p}\cdot {\bf p'}) 
+ 4  ({\bf p}\cdot {\bf p'})^2 
- \frac{4}{3} p^2 p'^2  \cr
B^d_{261}({\bf p},{\bf p}^{\prime})&=&  
  \frac{8}{3} p^4  
- \frac{16}{3} p^2 \;  ({\bf p}\cdot {\bf p'}) 
+ 4  ({\bf p}\cdot {\bf p'})^2 
- \frac{4}{3} p^2 p'^2 
\end{eqnarray}
The expressions for the functions $B^ d_{1k2}$, $k=1,6$, can be obtained from
the functions $B^ d_{2k1}$ by replacing ${\bf p} \leftrightarrow {\bf p'}$.

\section{Example of a chiral potential}
\label{appendixc}

For this particular example we will use the 
next-to-next-to-leading order (NNLO) chiral potential from Ref.~\cite{Ep05}. 

The leading-order (LO) NN potential in the two-nucleon center-of-mass system
(CMS) reads \cite{evgeny.report}:
\begin{equation}
V_{LO} = -\frac{1}{(2\pi)^3}  \frac{g_A^2}{4 F_\pi^2} \frac{ \bsigma_1 \cdot 
{\bf q} \, \bsigma_2 \cdot {\bf q}}{{\bf q} {}^2 +M_\pi^2} 
{\btau_1} \cdot {\btau_2}  + \frac{C_S}{(2\pi)^3} + \frac{C_T}{(2\pi)^3}
{\bsigma_1} \cdot {\bsigma_2}\,,
\label{eq:c1}
\end{equation}
where ${\bf q} = {\bf p} \, ' - {\bf p}$, and
$m_\pi$, $F_\pi$ and $g_A$ denote the pion mass, the pion decay constant 
and the nucleon axial coupling constants.
At next-to-leading order (NLO) a renormalization of the low energy constants 
(LECs) is required and the contribution from the Goldberger-Treiman discrepancy 
leads to a modified value of $g_A$.
The remaining contributions to the NN potential at this order are
\begin{eqnarray}
\label{VNLO}
V_{NLO} &=& - \frac{1}{(2\pi)^3} \frac{ {\btau}_1 \cdot {\btau}_2 }{384 \pi^2 F_\pi^4}\,
L^{\tilde \Lambda} (q) \, \Big[ 4m_\pi^2 (5g_A^4 - 4g_A^2 -1) 
+  {\bf q}\, ^2 (23g_A^4 - 10g_A^2 -1)
+ \frac{48 g_A^4 m_\pi^4}{4 m_\pi^2 + {\bf q} \, ^2} \Big] \nonumber \\
&-&  \frac{1}{(2\pi)^3}  \frac{3 g_A^4}{64 \pi^2 F_\pi^4} \,L^{\tilde \Lambda} (q)  \,
\bigg(  {\bsigma}_1 \cdot {\bf q}  \, {\bsigma}_2 \cdot {\bf q}   -  
{\bsigma}_1 \cdot{\bsigma}_2  \, {\bf q} \, ^2 \bigg) \cr 
&+& \frac{C_1}{(2\pi)^3} \, {\bf q}\,^2 + \frac{C_2}{(2\pi)^3} \, {\bf k}^2 +
( \frac{C_3}{(2\pi)^3} \, {\bf q}\,^2 + \frac{C_4}{(2\pi)^3} \, {\bf k}^2 ) \,  {\bsigma}_1 \cdot
{\bsigma}_2 \nonumber \\
&+& \frac{C_5}{(2\pi)^3}\, \frac{i}{2} \, ( {\bsigma}_1 + {\bsigma}_2) \cdot  {\bf q} \times
{\bf k}
 + \frac{C_6}{(2\pi)^3} \, {\bf q}\cdot {\bsigma}_1 \, {\bf q}\cdot {\bsigma}_2
+ \frac{C_7}{(2\pi)^3} \, {\bf k}\cdot {\bsigma}_1 \, {\bf k}\cdot {\bsigma}_2 \,,
\label{eq:c2}
\end{eqnarray}
where  
$q \equiv | {\bf q} \, |$ and 
$ {\bf k} = \frac12 \left( {\bf p} \, ' + {\bf p} \, \right)$.
The loop function
$L^{\tilde \Lambda} (q)$ is defined in the spectral function regularization (SFR)
as~\cite{Ep05}
\begin{equation}
L^{\tilde \Lambda} (q) = \theta (\tilde \Lambda - 2 m_\pi ) \, \frac{\omega}{2 q} \,
\ln \frac{\tilde \Lambda^2 \omega^2 + q^2 s^2 + 2 \tilde \Lambda q
\omega s}{4 m_\pi^2 ( \tilde \Lambda^2 + q^2)} \,,
\label{eq:c3}
\end{equation}
with the following abbreviations:
$\omega = \sqrt{4 m_\pi^2 + {\bf q}\, ^2}$ and
$s = \sqrt{\tilde \Lambda^2 - 4 m_\pi^2}$. Here, $\tilde \Lambda$ denotes the
ultraviolet cutoff in the mass spectrum of the two-pion-exchange potential. 

The contributions at NNLO again lead to the renormalization and/or redefinition of the LECs $C_S$, $C_T$, $C_1, \dots C_7$.  
The only new momentum dependence is due to the 
following terms:
\begin{eqnarray}
V_{NNLO} &=&  - \frac{1}{(2\pi)^3} \frac{3g_A^2}{16\pi F_\pi^4}  \big( 2m_\pi^2(2c_1 -c_3)
-c_3 {\bf q} \, ^2 \big)
 (2m_\pi^2+{\bf q} \, ^2) A^{\tilde \Lambda} (q) \nonumber \\
& -& \frac{1}{(2\pi)^3} \frac{g_A^2 c_4 }{32\pi F_\pi^4} \, {\btau}_1 \cdot {\btau}_2 \,  
(4m_\pi^2 + q^2) A^{\tilde \Lambda}(q)
\big( {\bsigma}_1 \cdot {\bf q}\, {\bsigma}_2 \cdot {\bf q}\,
-{\bf q} \, ^2 \, {\bsigma}_1 \cdot{\bsigma}_2 \big)\,,
\label{eq:c4}
\end{eqnarray}
where $c_1$, $c_3$, $c_4$ are new $\pi$N LECs
and the loop function $A^{\tilde \Lambda} (q)$ is given by
\begin{equation}
A^{\tilde \Lambda} (q) = \theta (\tilde \Lambda - 2 m_\pi ) \, \frac{1}{2 q} \,
\arctan \frac{q ( \tilde \Lambda - 2 m_\pi )}{q^2 + 2 \tilde \Lambda m_\pi}\,.
\label{eq:c5} 
\end{equation}
The expressions of the potential given in Eqs.~(\ref{eq:c1}), (\ref{eq:c2}),
and (\ref{eq:c4}) show that this potential can be readily expressed in the
operators $w_j$, $j=1,6$ of Eq.~(\ref{eq:2.2}). 
 
The chiral potential we consider in this example is the sum of 
\begin{equation}
V  \equiv  V_{LO} + V_{NLO} + V_{NNLO} .
\end{equation}
It requires regularization when inserted into the Lippmann-Schwinger
equation, which is achieved by introducing a regulated potential of the
form
\begin{equation}
V_{reg} ( {\bf p}^{\ \prime} , {\bf p} \, )  \equiv 
e^{-\left(p^{\prime \,4} /\Lambda^4   \right)} \,
V ( {\bf p}^{\ \prime} , {\bf p} \, ) \, 
e^{-\left(p^{4} /\Lambda^4 \right)} \, ,
\end{equation}
with the cut-off parameter $\Lambda$.

\section{Scalar functions for the Bonn B potential}
\label{appendixd}

For the convenience of the reader we give the expressions for the Bonn B
potential
from Ref.~\cite{machl} in a form which is more suited for our three-dimensional calculations. 
The expressions for the exchange of pseudo-scalar ($ps$), scalar ($s$), and
vector ($v$) mesons are given by
\begin{eqnarray}
V_{ps}({\bf p}',{\bf p}) & = & \frac{g_{ps}^{2}}{(2\pi
)^{3}4m^{2}}\sqrt{\frac{m}{E'}}
\sqrt{\frac{m}{E}}\frac{F^{2}_{ps}[({\bf p}'-{\bf p})^2]}{({\bf p}'-{\bf p})^{2}
+m_{ps}^{2}}\frac{O_{ps}}{W'W}\cr
V_{s}({\bf p}',{\bf p}) & = & \frac{g_{s}^{2}}{(2\pi
)^{3}4m^{2}}\sqrt{\frac{m}{E'}}\sqrt{\frac{m}{E}}\frac{F^{2}_{s}[({\bf
p}'-{\bf p})^2]}{({\bf p}'-{\bf p})^{2}+m_{s}^{2}}\frac{O_{s}}{W'W} \cr
V_{v}({\bf p}',{\bf p}) & = & \frac{1}{(2\pi
)^{3}4m^{2}}\sqrt{\frac{m}{E'}}\sqrt{\frac{m}{E}}\frac{F^{2}_{v}[({\bf p}'-{\bf p})^2]}{({\bf p}'-{\bf
p})^{2}+m_{v}^{2}} \frac{\left( g_{v}^{2} O_{vv}+2g_{v}f_{v} O_{vt}+ f_{v}^{2} O_{tt}\right) }{W'W} ,
\qquad
\label{eq:d1}
\end{eqnarray}
where $m_{\alpha}$ are the masses of the exchanged mesons, $m$ the nucleon
mass,  $E=\sqrt{m^2+{\bf p}^2}$ and $ W = m + E $. 
The crucial quantities are the operators $O_{ps}$, $O_{s}$, $O_{vv}$, $ O_{vt}$ 
and $O_{tt}$, which are given in 
terms of the Dirac spinors as
\begin{equation}
O_{ps} = 4 m^2 W' W \, \bar{u}({\bf p}')\gamma^5 u({\bf p})\bar{u}(-{\bf p}')\gamma^5 u(-{\bf p}) ,
\label{eq:d2}
\end{equation}
\begin{equation}
O_{s} = -4 m^2 W' W \, \bar{u}({\bf p}') u({\bf p})\, \bar{u}(-{\bf p}') u(-{\bf p}) ,
\label{eq:d3}
\end{equation}
\begin{equation}
O_{vv} = 4 m^2 W' W \, \bar{u}({\bf p}')\gamma^\mu u({\bf p})\bar{u}(-{\bf p}')\gamma_\mu u(-{\bf p}) ,
\label{eq:d4}
\end{equation}
\begin{eqnarray}
O_{vt} &=& m W' W \, \Big\{ 4m\, \bar{u}({\bf p}')\gamma^\mu
u({\bf p})\bar{u}(-{\bf p}')\gamma_\mu u(-{\bf p})  \cr
&-&\bar{u}({\bf p}')\gamma^\mu u({\bf p})\bar{u}(-{\bf p}')
\bigg[(E'-E)(g_\mu^0-\gamma_\mu \gamma^0)+(p_{2}+p_{2}')_\mu \bigg] 
u(-{\bf p})\cr
&-&\bar{u}({\bf p}') \bigg[(E'-E)(g^{0\mu }-\gamma ^\mu \gamma^0)+
(p_{1}+p_{1}')^{\mu } \bigg] u({\bf p})\bar{u}(-{\bf p}')\gamma_\mu u(-{\bf
p}) \Big\} ,
\label{eq:d5}
\end{eqnarray}
\begin{eqnarray}
O_{tt}& =& 
W' W \,
\Big\{ 4m^{2}\, \bar{u}({\bf p}')\gamma ^{\mu }u({\bf p})\bar{u}(-{\bf
p}')\gamma _{\mu }u(-{\bf p}) \cr
&-& 2m\, \bar{u}({\bf p}')\gamma ^{\mu }u({\bf p})\bar{u}(-{\bf p}') 
\bigg[(E'-E)(g_{\mu }^{0}-\gamma _{\mu }\gamma ^{0})+(p_{2}+p_{2}')_{\mu
}\bigg] u(-{\bf p}) \cr
&-&2m\, \bar{u}({\bf p}')\bigg[(E'-E)(g^{0\mu }-\gamma ^{\mu }\gamma
^{0})+(p_{1}+p_{1}')^{\mu }\bigg]
u({\bf p})\bar{u}(-{\bf p}')\gamma _{\mu }u(-{\bf p}) \cr
&+ &\bar{u}({\bf p}')\bigg[(E'-E)(g^{0\mu }-\gamma ^{\mu }\gamma
^{0})+(p_{1}+p_{1}')^{\mu }\bigg] u({\bf p}) \cr
 &\times & \bar{u}(-{\bf p}') \bigg[(E'-E)(g_{\mu }^{0}-\gamma _{\mu }\gamma
^{0})+(p_{2}+p_{2}')_{\mu } \bigg] u(-{\bf p})\Big\} 
\label{eq:d6}
\end{eqnarray}
with $ (p_1 + {p_1} ')^ {\mu} = ( E + E ', {\bf p} + {\bf p'})$ and
 $(p_2 + p_2 ')^ {\mu} = ( E + E ', - {\bf p} - {\bf p'})$.

These operators act in the spin spaces of nucleons 1 and 2:
the bilinear forms built with
$ \bar{u}({\bf p}') \dots u({\bf p})$ contain $ {\bsigma}_1 $ 
as acting in the spin space of nucleon 1
and the bilinear forms with $ \bar{u}(-{\bf p}')\dots u(-{\bf p})$
contain $ {\bsigma}_2 $ and act in the spin space of nucleon 2.
The spinors $u({\bf q})$ are normalized according to the definitions given in
Ref.~\cite{Bjorken} and explicitly given as
\begin{eqnarray}
u({\bf q}) = \sqrt{\frac{E+m}{2m}} \left( \begin{array}{c} 1 
\\ \frac{ \bsigma \cdot {\bf q}}{E+m} \end{array}\right) .
\label{eq:d7}
\end{eqnarray}
Each vertex is multiplied with a form factor 
\begin{equation}
F^2_{\alpha }[({\bf p}'-{\bf p})^2] = \left( \frac{\Lambda ^{2}_{\alpha }-m_{\alpha }^{2}}{\Lambda
^{2}_{\alpha }+({\bf p}'-{\bf p})^{2}}\right) ^{2n} .
\label{eq:d8}
\end{equation}
where the values of $n$ and the cutoff parameters $\Lambda _{\alpha }$ 
are given in Table~\ref{tbonnb}.

Note that for the three iso-vector mesons ($\pi$, $\delta$ and $\rho$)
contributing to the Bonn B potential,
expressions (D2)--(D6) are additionally multiplied by the isospin 
factor $ {\btau}(1) \cdot {\btau}(2)$.

In Ref.~\cite{imam} this potential was presented in a different operator 
form. However, in that work one of the six operators  was chosen to be
$~{\bsigma}_1 \cdot ( {\bf p} + {\bf p'}) \;  {\bsigma}_2 \cdot ( {\bf p'} -
{\bf p}) 
 + {\bsigma}_1 \cdot  ( {\bf p'} - {\bf p})\;  {\bsigma}_2 \cdot ( {\bf p} +
{\bf p'})~$, 
which is an operator that violates time reversal invariance.  In practice, this
operator is always multiplied with the term
 $( p'^2 - p^2)$, which also violates time reversal invariance. Therefore, the
entire term is invariant as it
should be. In principle it is not desirable to work with symmetry violating
operators, thus we prefer to use the operators from Eq.~(\ref{eq:2.2}) and
rewrite
\begin{eqnarray}
{\bsigma}_1\cdot ({\bf p} + {\bf p'}) \; {\bsigma}_2\cdot({\bf p'}-{\bf p}) 
&+& {\bsigma}_1 \cdot  ( {\bf p'} - {\bf p}) \;  {\bsigma}_2 \cdot ( {\bf p} +
{\bf p'}) = \cr
\frac{-4 ({\bf p}\times{\bf p'})^2}{{ p'}^2 -p^2 } \; {\bsigma}_1
\cdot {\bsigma}_2  &+& 
\frac{({\bf p}-{\bf p'})^2}{{ p'}^2- p^2} {\bsigma}_1\cdot
({\bf p}+{\bf p'}) \; {\bsigma}_2 \cdot ({\bf p}+{\bf p'}) \cr
 + \frac{  ( {\bf p} + {\bf p'})^2 }{  { p'}^ 2  - p^ 2 }
{\bsigma}_1 \cdot ( {\bf p} - {\bf p'}) \; {\bsigma}_2 \cdot ( {\bf p} - {\bf
p'}) & +& 
\frac{ 4 }{  { p'}^ 2  - p^ 2 }
{\bsigma}_1 \cdot ( {\bf p} \times {\bf p'}) \; {\bsigma}_2 \cdot ( {\bf p}
\times {\bf p'}) ,
\label{eq:d9}
\end{eqnarray}
which is an identity for $({\bf p}+{\bf p'})({\bf p'}-{\bf p})= p'^2-p^2 \ne 0 $.
Inserting Eq.~(\ref{eq:d9})
into the expressions given in \cite{imam} cancels the factor $p'^2-p^2$ 
and one obtains the following expressions for the operators
$O_{\alpha}$ from Eqs.~(\ref{eq:d2})-(\ref{eq:d6}) in terms of the
operators 
$ w_j \equiv  w_j( {\bsigma}_1, {\bsigma}_2, {\bf p'}, {\bf p})$ from
Eq.~(\ref{eq:2.2}):
\begin{eqnarray}
O_{ps} & = & \left(1 + \frac{2m}{E'+E}\right) \bigg[ ({\bf p}' \cdot {\bf p})^2 -
p'^2 p^2 \bigg] w_2 
   + \left(1 + \frac{2m}{E'+E}\right) w_4 \cr
&& + \frac{1}{4}\Biggl\lbrace - (W'-W)^{2} 
   + \left(1 + \frac{2m}{E'+E}\right) [p'^2 + p^2 - 2({\bf p}' \cdot {\bf p})] \Biggr\rbrace w_5 \cr
&& + \frac{1}{4}\Biggl\lbrace - (W'+W)^2 
   + \left(1 + \frac{2m}{E'+E}\right) [p'^2 + p^2 + 2({\bf p}' \cdot {\bf p})] \Biggr\rbrace w_6 
\label{eq:d.10}
\end{eqnarray}
\begin{eqnarray}
O_{s} & = & -[W'W-({\bf p}' \cdot {\bf p})]^2 w_1 
            -[W'W-({\bf p}' \cdot {\bf p})] w_3 + w_4 
\label{eq:d.11}
\end{eqnarray}
\begin{eqnarray}
O_{vv} & = & \left\lbrace [W'W+({\bf p}' \cdot {\bf p})]^2 + W'^{2}p^{2}+W^{2}p'^{2}+2W'W({\bf p}' \cdot {\bf p}) \right\rbrace w_1 \cr
&& + \Biggl\lbrace - \frac{1}{2}\left(W'^{2} + W^{2}\right)\left(p'^{2} + p^{2}\right) + 2W'W({\bf p}' \cdot {\bf p}) \cr 
&& + \frac{1}{2}\left(1 + \frac{2m}{E'+E}\right)\left[ p'^{4} + p^{4} - 2({\bf p}' \cdot {\bf p})^2 \right] \Biggr\rbrace w_2 \cr
&& -[3W'W+({\bf p}' \cdot {\bf p})] w_3 
   - \left(2 + \frac{2m}{E'+E}\right) w_4 \cr
&& - \frac{1}{4} \left\lbrace - (W'-W)^{2} 
   + \left(1 + \frac{2m}{E'+E}\right) [p'^2 + p^2 - 2({\bf p}' \cdot {\bf p})] \right\rbrace w_5 \cr
&& - \frac{1}{4} \left\lbrace - (W'+W)^2 
   + \left(1 + \frac{2m}{E'+E}\right)[p'^2 + p^2 + 2({\bf p}' \cdot {\bf p})] \right\rbrace w_6 
\end{eqnarray}
\begin{eqnarray}
O_{vt} & = & \left\lbrace W'^{2}p^{2}+W^{2}p'^{2}-\frac{W'-W}{2m}\left(W'^{2}p^{2}-W^{2}p'^{2}\right) \right. \cr 
 &  & + \left. 2W'W\left[2W'W+({\bf p}' \cdot {\bf p})\right]-\frac{W'+W}{m}\left[W'^{2}W^{2}-({\bf p}' \cdot {\bf p})^{2}\right] \right\rbrace w_1 \cr
 &  & +\left\{ 2W'W({\bf p}' \cdot {\bf p}) - \frac{1}{2}\left(W'^{2}+W^{2}\right)\left(p'^{2}+p^{2}\right) \right. \cr
&& + \frac{1}{2}\left(1 + \frac{2m}{E'+E}\right)\left[ p'^{4} +p^{4} - 2({\bf p}' \cdot {\bf p})^2 \right] \cr 
 &  & - \left. \frac{1}{2m(E'+E)}  \left[ W'^{2}p^4+W^{2}p'^4 - \left(W'^{2}+W^{2}\right)({\bf p}' \cdot {\bf p})^2 \right] \right\} w_2 \cr
 &  & -\left[ 2W'W +\frac{W'+W}{m}({\bf p}' \cdot {\bf p})\right] w_3 \cr 
 &  & +\left\{ -\frac{W'+W}{m} +\frac{1}{2m(E' + E)}\left[W'^{2}+W^{2}-2m(W'+W)\right] \right\} w_4 \cr
 &  & -\frac{1}{4} \left\{ -(W'-W)^{2} + \left(1 + \frac{2m}{E'+E}\right)[p'^2 + p^2 - 2({\bf p}' \cdot {\bf p})]\right. \cr 
 &  & - \left. \frac{1}{m(E' + E)} [W'^{2}p^2 + W^{2}p'^2 - \left(W'^{2}+W^{2}\right)({\bf p}' \cdot {\bf p})] \right\} w_5 \cr
 &  & -\frac{1}{4} \left\{ -(W'+W)^{2}+\left(1 + \frac{2m}{E'+E}\right)[p'^2 + p^2 + 2({\bf p}' \cdot {\bf p})] \right. \cr 
 &  & - \left. \frac{1}{m(E' + E)}[W'^{2}p^2+W^{2}p'^2 + \left(W'^{2}+W^{2}\right)({\bf p}' \cdot {\bf p})] \right\} w_6 
\label{eq:d.12}
\end{eqnarray}
\begin{eqnarray}
O_{tt}& = &\Biggl\{ \left[W'W + ({\bf p}' \cdot {\bf p})\right]^2 
      + 2 \left( 2 -\frac{W'+W}{m} \right) \left[W'^2 W^2 - ({\bf p}' \cdot {\bf p})^2\right] \cr
 &  & + \left\lbrace 3 + \frac{3[W'W-m(W'+W)]+({\bf p}' \cdot {\bf p})}{2m^2} \right\rbrace \left[W' W - ({\bf p}' \cdot {\bf p}) \right]^2 \cr
 &  & +\left[1+\frac{(W'-W)^2}{4m^2}\right] \left(W'^{2}p^{2}+W^{2}p'^{2}\right) + 2 \left[ 1 -\frac{(W'-W)^2}{4m^2} \right]W'W({\bf p}' \cdot {\bf p}) \cr
 &  & -\frac{W'-W}{m} \left(W'^{2}p^{2}-W^{2}p'^{2}\right) \Biggr\} w_1 \cr 
 &  & +\Biggl\{ 2\left[1-\frac{(W'-W)^{2}}{4m^2}\right]W'W({\bf p}' \cdot {\bf p}) -\left[1+\frac{(W'-W)^2}{4m^2}\right]\left(1 + \frac{2m}{E'+E}\right) ({\bf p}' \cdot {\bf p})^2\cr
 &  & -\left[1+\frac{(W'-W)^2}{4m^2}\right] \left(\frac{m^2 +E'E}{E'+E}+m\right)(W'p^2 + Wp'^2) \cr
 &  & -\frac{1}{m(E' + E)}\left[ W'^{2}p^4 +W^{2}p'^4 -\left(W'^{2}+ W^2\right)({\bf p}' \cdot {\bf p})^2 \right] \Biggr\} w_2 \cr
 &  & +\Biggl\lbrace - W'W - ({\bf p}' \cdot {\bf p}) + 2 \left( 2 - \frac{W'+W}{m} \right) ({\bf p}' \cdot {\bf p}) - 2 \left[1 - \frac{(W'-W)^2}{4m^2} \right] W'W \cr
 &  & + \left\lbrace 3 + \frac{3[W'W - m(W'+W)]+({\bf p}' \cdot {\bf p})}{2m^2} \right\rbrace [W'W - ({\bf p}' \cdot {\bf p})] \Biggr\rbrace w_3  \cr
 &  & -\left\{ 4 + \frac{3[W'W-m(W'+W)]+({\bf p}' \cdot {\bf p})}{2m^{2}} - 2 \left(2 - \frac{W'+W}{m}\right) \right. \cr 
 &  & + \left[1+\frac{(W'-W)^2}{4m^2}\right]\left(1 + \frac{2m}{E'+E}\right) 
      - \frac{1}{m(E' + E)}\left(W'^{2}+W^2\right) \Biggr\} w_4 \cr
 &  & -\frac{1}{2} \Biggl\{ \left[1-\frac {(W'-W)^{2}}{4m^2}\right]W'W -\frac{1}{m(E'+E)} [W'^2p^2+W^2p'^2 - \left(W'^{2}+W^2\right)({\bf p}' \cdot {\bf p})] \cr 
 &  & +\frac{1}{2}\left[1+\frac{(W'-W)^2}{4m^2}\right] \left[ \left(1 + \frac{2m}{E'+E}\right) [p'^2 + p^2 - 2({\bf p}' \cdot {\bf p})] - W'^2 - W^2 \right] \Biggr\}  w_5 \cr
 &  & -\frac{1}{2}\Biggl\{ -\left[1-\frac{(W'-W)^{2}}{4m^2}\right]W'W 
      - \frac{1}{m(E' + E)}\left[W'^2p^2+W^2p'^2 + (W'^2+W^2)({\bf p}' \cdot {\bf p})\right] \cr 
 &  & + \frac{1}{2}\left[1+\frac{(W'-W)^2}{4m^2}\right]\left[\left(1 +
\frac{2m}{E'+E}\right)[p'^2 + p^2 + 2({\bf p}' \cdot {\bf p})] -W'^2-W^2\right] \Biggr\} w_6 .
\label{eq:d.13}
\end{eqnarray}
The values of the parameters 
are given in Table~\ref{nntabobep}.


\begin{thebibliography}{99}

\bibitem{wgrep}W. Gl\"ockle, H. Wita{\l}a, D. H\"uber, H. Kamada, J. Golak,
Phys. Rep. 274, 107 (1996).

\bibitem{Nogga:2000uu}
  A.~Nogga, H.~Kamada and W.~Gl\"ockle,
  Phys. Rev. Lett.  {\bf 85}, 944 (2000).

\bibitem{Elster:1998qv}
  Ch.~Elster, W.~Schadow, A.~Nogga and W.~Gl\"ockle,
  Few Body Syst.\  {\bf 27}, 83 (1999).

\bibitem{Liu:2004tv}
  H.~Liu, Ch.~Elster and W.~Gl\"ockle,
  Phys. Rev.  C {\bf 72}, 054003 (2005).


\bibitem{imam} I. Fachruddin, Ch. Elster, W. Gl\"ockle,
Phys. Rev. C{\bf 63}, 054003 (2001).

\bibitem{Bayegan:2007ih}
  S.~Bayegan, M.~R.~Hadizadeh and M.~Harzchi,
  Phys.\ Rev.\  C {\bf 77}, 064005 (2008).

\bibitem{Ramalho:2006jk}
  G.~Ramalho, A.~Arriaga and M.~T.~Pena,
  Few Body Syst.\  {\bf 39}, 123 (2006).

\bibitem{Caia:2003ke}
  G.~Caia, V.~Pascalutsa and L.~E.~Wright,
  Phys.\ Rev.\  C {\bf 69}, 034003 (2004).

\bibitem{RodriguezGallardo:2007dc}
  M.~Rodriguez-Gallardo, A.~Deltuva, E.~Cravo, R.~Crespo and A.~C.~Fonseca,
  Phys. Rev. C {\bf 78}, 034602 (2008).



\bibitem{2N3N} W. Gl\"ockle, Ch. Elster, J. Golak, R. Skibi\'nski,
H. Wita{\l}a, H. Kamada, arXiv:0906.0321, Few-Body Syst.
DOI 10.1007/s00601-009-0064-1. 

\bibitem{3Nscatt} W. Gl\"ockle, I. Fachruddin, Ch. Elster, J. Golak, 
R. Skibi\'nski, H. Wita{\l}a, arXiv:0910.1177, 
accepted for publication in EPJ {\bf A}.

\bibitem{Ep05} E.~Epelbaum, W.~Gl\"ockle, U.-G.~Mei{\ss}ner,
Nucl. Phys. A \textbf{747}, 362 (2005).

\bibitem{evgeny.report} E. Epelbaum, 
Prog. Part. Nucl. Phys. {\bf 57}, 654 (2006).

\bibitem{newer.report} E.~Epelbaum, H.~W.~Hammer, and U.-G.~Mei{\ss}ner, 
arXiv:0811.1338.

\bibitem{machl} R. Machleidt, Adv. Nucl. Phys. {\bf 19}, 189 (1989).

\bibitem{ISloan} I. Sloan, Comp. Phys. {\bf 3}, 332 (1968).

\bibitem{wolfenstein} L. Wolfenstein, Phys. Rev. {\bf 96}, 1654 (1954).

\bibitem{book} W. Gl\"ockle, {\em The Quantum Mechanical Few-Body Problem},
Springer-Verlag, Berlin-Heidelberg, (1983).

\bibitem{stapp} M. H. McGregor, M. J. Moravcsik, H. P. Stapp, Annu. Rev. Nucl. Sci. {\bf 10}, 291 ( 1960).

\bibitem{numrec} W. Press, B. Flannery, S. Teukolsky, and W. Vetterling, {\it Numerical Recipes}, Cambridge University 
Press, 1989.

\bibitem{math} Wolfram Research, Inc., {\em Mathematica$^\copyright$}, 
Version 7.0, Champaign, IL (2008).

\bibitem{tinglin}
T.~Lin, C.~Elster, W.~N.~Polyzou, H.~Wita\l a and W.~Gl\"ockle,
Phys. Rev. C{\bf 78}, 024002 (2008).

\bibitem{Bjorken} J.D. Bjorken, S.D. Drell, {\it Relativistic Quantum
Mechanics}, McGraw-Hill Science/Engineering/Math, 1998.

\end{thebibliography}
\end{document}